\documentclass[%
reprint,
superscriptaddress,
showpacs,
hidelinks,
amsmath,amssymb,
aps,
prb,
a4paper
]{revtex4-1}
\usepackage{graphicx,epstopdf}
\usepackage{placeins}
\usepackage{tikz}
\usepackage{tikz-3dplot}
\usetikzlibrary{calc,math}
\usetikzlibrary{intersections}
\usetikzlibrary{positioning}
\usetikzlibrary{arrows,decorations.markings}
\usepackage{pgfplots}
\pgfplotsset{compat=newest}
\usepgflibrary{shapes.geometric}
\usepackage{mathrsfs}
\usepackage{bm}
\usepackage{esint}
\usepackage{siunitx}
\usepackage{physics}
\usepackage{xurl}
\usepackage{hyperref}
\usepackage[hyphenbreaks]{breakurl}

\usepackage[normalem]{ulem}
\usepackage{cancel}

\usepackage{xcolor} 

\begin{document}
\title{Flux-periodic oscillations in proximitized core-shell nanowires}
\author{Kristjan Ottar Klausen}
\email{kristjan19@ru.is}
\affiliation{Department of Engineering, Reykjavik University, Menntavegur 1, IS-101 Reykjavik, Iceland.}
\author{Anna Sitek}
\affiliation{Department of Theoretical Physics,
	Wroclaw University of Science and Technology,
	Wybrze{\.z}e Wyspia{\'n}skiego 27, 50-370 Wroclaw, Poland.}
\author{Sigurdur I.\ Erlingsson}
\affiliation{Department of Engineering, Reykjavik University, Menntavegur 1, IS-101 Reykjavik, Iceland.}
\author{Andrei Manolescu}
\affiliation{Department of Engineering, Reykjavik University, Menntavegur 1, IS-101 Reykjavik, Iceland.}
\begin{abstract}
Flux-periodic oscillations of the superconducting gap in proximitized core-shell nanowires are explored. Periodicity of oscillations in the energy spectrum of a cylindrical nanowire is compared with nanowires having hexagonal and square cross-section geometry, along with the effects of Zeeman and Rashba spin-orbit interaction. A transition between h/e and h/2e periodicity is found and shown to be dependent on the chemical potential, with correspondence to degeneracy points of the angular momentum quantum number. For a thin shell of a square nanowire, solely h/e periodicity is found in the infinite wire spectrum and shown to result from energy separation between the lowest groups of excited states.
\end{abstract}
\maketitle

\section{INTRODUCTION}
In the past decades, periodicity of magneto-oscillations has been broadly discussed in relation to doubly connected mesoscopic systems, both normal conducting\cite{Zhu_95} and superconducting \cite{Yang_61}. In normal conducting rings, the periodicity is commonly $h/e$, whereas in conventional superconductors it is found in terms of the flux quantum, $h/2e$.
 Superconducting cylinders exhibit the Little-Parks effect \cite{LittleParks} when exposed to an external axial magnetic field, where oscillations of the superconducting transition temperature occur as a function of the superconducting flux quantum. The factor of $2e$ is interpreted to be a signature of the charge pairing in Cooper pairs, although alternative mechanisms have been suggested \cite{Fan_1999}.
The effect can be found in both solid core and hollow core full-shell nanowires where the wire is fully coated by a superconducting shell \cite{evenodd}, and superconductivity is induced in the coated wire by the proximity effect \cite{2019ProxCoreShell}.

Core-shell nanowires are radial heterojunctions, where a nanowire core is enveloped by layers of different materials. 
By combining a resistive core with a conductive shell, a tubular conductor can be formed in a core-shell nanowire \cite{Blomers_13, Pistol_2008}. Due to crystallographic properties, nanowire cross-sections are most commonly polygonal, where hexagonal\cite{Rieger2012}, square\cite{Fan_06}, and triangular\cite{TriHex} cross-sections have been fabricated. 
The polygonal corners cause electrons to localize unevenly along the circumference of the shell. Specifically, electrons with low energy tend to accumulate near the sharp edges, while those with higher energy localize at the sides \cite{Ferrari09b, Sitek15}. For thin shells, the corner localization dominates such that each edge behaves effectively as single wire, which has been shown to allow for hosting multiple pairs of Majorana zero modes in a single wire system \cite{Andrei}.
In particular, InSb shells\cite{Zellekens_2020} provide high electron mobility \cite{G_l_2015}, a large g-factor and strong spin-orbit interaction \cite{Wojcik2019,van_Weperen_2015}, which are favorable properties for the realization of Majorana zero modes \cite{Alicea_2012,Stanescu2011, Zhang2019}.

 Flux-periodicity is considered to be an experimental signature that can probe the underlying pairing symmetry of superconducting systems \cite{Tsuei_1996}. 
  Periodicity of $h/2e$ is however found in magnetoresistance oscillations of disordered normal conducting metal rings \cite{Fourcade_86,Qiming_86,Pannatier_85,AAS_81} and in systems with a spin gap \cite{Seidel_05}. Conversely, $h/e$ periodicity has been shown to emerge in magneto-oscillations of  the current density in superconducting square loops\cite{Zha_2012,Loder2008}.
 Furthermore, fractionalized flux-quantum periodicity has been measured in disordered superconducting cylinders, as well as spin-triplet superconductors \cite{Xu_2020,Yasui_2017,Cai_13,Nazarov_89,Geshkenbein_87}. 
 Triplet pairing has been linked to the formation of half-quantum vortices in p-wave superconductors \cite{Ivanov_2001}, Bose-Einstein condensates \cite{Seo_2015,Manni_2012} and Superfluid Helium-III \cite{Autti_2016,Volovik_1987}. Flux-periodicity has been hypothesized to derive from off-diagonal long range order \cite{Nieh_95,Yang_1962} where multi-component order parameters lead to fractionalized vortices \cite{Rampp_2022}. Vortices with an arbitrary fraction of the flux quantum have been shown to exist in two-gap superconductors \cite{Babaev_2002}. Fractional flux quanta have also been demonstrated for looped solenoids \cite{Melo_96}. 
 Clearly, various factors can influence flux-periodicity in a wide range of mesoscopic systems.
 A common thread is mixed pairing leading to nonintegral flux values \cite{Scwartz_65,Schwartz_64}.
 
Semiconductor nanowires with proximity induced superconductivity incorporate both spin-dependent interactions and superconductor pairing.\cite{FS_Marcus22,FS_Marcus20} In order to clarify the effects of cross-section geometry, Zeeman splitting and spin-orbit interaction on flux-periodicity, we explore magneto-oscillations in the energy spectrum of a proximitized core-shell nanowire in an external axial magnetic field, modeled by the Bogoliubov-de Gennes (BdG) Hamiltonian \cite{Jianxin,Bogoliubov:1958km}.

The paper is structured as follows. In Sect.\ \ref{FP}, the Little-Parks effect is introduced and contrasted with flux-periodic oscillations in normal conducting rings and cylinders using a simple model. Sect.\ \ref{MM} details the nanowire model, basis and parameters. In Sect.\ \ref{CYL}, flux-periodic oscillations in a cylindrical wire are compared for the cases of vanishing/nonvanishing Zeeman and spin-orbit interaction, along with energy spectra and dispersions for both the normal conducting and superconducting wire. Sect.\ \ref{POLY} consists of an analogous treatment for wires with hexagonal and square cross-sections, along with a detailed discussion of the effect of the chemical potential on periodicity for the hexagonal case. Lastly, conclusions are summarized in Sect.\ \ref{CONCL}.

\section{FLUX-PERIODIC OSCILLATIONS}
\label{FP}
The Little-Park effect is attributed to supercurrents induced within the walls of a thin superconducting cylinder\cite{SanJose_2022}.
The expression for the supercurrent in the semi-classical Ginzburg-Landau theory is the following, \cite{GL}
\begin{equation}
	\mathbf{J}_s = \frac{-i\hbar e^*}{2m^*} \left(\psi^\dagger \bm{\nabla} \psi - \psi \bm{\nabla} \psi^\dagger \right) - \frac{e^{*2}}{m^*} \, \abs{\psi}^2\mathbf{A} \ .
	\label{JsGL2}
\end{equation}
Rewriting the supercurrent with
$\psi = \abs{\psi}e^{i\phi}$ and integrating around a closed contour
gives \cite{FetterW}
\begin{equation}
	\oint_C \mathbf{A} \cdot d\bm{\ell} + \frac{m^*}{e^{*2}} \oint_C \abs{\psi}^{-2} \mathbf{J_S}\cdot d\bm{\ell} = \frac{\hbar}{e^*}\oint_C \bm{\nabla} \phi \cdot d\bm{\ell} \ ,
	\label{fluxoidint}
\end{equation}
where $e^*$ and $m^*$ are effective charge and mass.
The right hand side integral must be an integer multiple $n \in \mathbb{N}$ of 2$\pi$ if $\psi$ is to be single valued.  In the case of Cooper pairing, $e^*=2e$, leading to the superconducting flux quantum
\begin{equation}
	\Phi_0^{SC}= \frac{h}{2e} \ .
\end{equation}
The left hand side of Eq.\ \eqref{fluxoidint} can be rewritten in terms of the magnetic flux and supercurrent velocity \cite{Tinkham} to obtain
\begin{equation}
	\oiint_S \mathbf{B} \cdot d\mathbf{S} + \frac{m^*}{2e} \oint \mathbf{v}_s = n \cdot \Phi_0^{SC},
\end{equation}
which can be rewritten as
\begin{equation}
	v_s= \frac{\hbar}{rm^*} \left(n - \frac{\Phi_B}{\Phi_0^{SC}}\right).
\end{equation}
Since the critical temperature and coherence length are proportional to the square of the supercurrent velocity \cite{Tinkham}, both become periodic functions of the magnetic flux with periodicity of $\Phi_0$ \cite{Golubov03}.

In normal conducting rings and cylinders, flux periodicity of the energy spectra occurs as well with periodicity of $\Phi_0^N=h/e$.
Consider a ring in the xy-plane in a uniform magnetic field $\mathbf{B}=B \, \mathbf{e}_z$ perpendicular to the plane of the ring.
The total magnetic flux through the area enclosed by the radius of the ring, $r$, is 
\begin{equation}
	\Phi_B=\iint_S \mathbf{B}\cdot d\mathbf{S}= \pi r^2 B,
\end{equation}
which can equally be expressed in terms of the magnetic vector potential using Green's curl theorem, working in cylindrical coordinates $(r,\theta,z)$,
\begin{equation}
	\Phi_B= \oint_s \mathbf{A} \cdot d\mathbb{\ell} = 2\pi r A_{\theta} .
\end{equation}	
The Hamiltonian of the system is given by
\begin{equation}
	H_t= \frac{1}{2m_e} (\mathbf{p}-e\mathbf{A})^2 = \frac{1}{2m_e}\left(-\frac{i\hbar}{r} \frac{\partial}{\partial\theta} - \frac{e\Phi_B}{2\pi r}\right)^2.
	\label{basicHam}
\end{equation}
The effective mass is $m_e=m_x\cdot m_0$ where $m_0$ is the electron rest mass and the parameter $m_x$ is material dependent.
Solving the time-independent Schr\"odinger equation $\hat{H}\psi=E\psi$ with the eigenfunctions $\psi=e^{il\theta}$ results in the energy spectrum
\begin{equation}
	E_l=\frac{1}{2m_e} \left(\frac{\hbar}{r} l - \frac{e\Phi_B}{2\pi r} \right)^2  \ .
\end{equation}
Rewriting in terms of the normal flux quantum, $\Phi_0^N=\frac{2\pi \hbar }{e}=\frac{h}{e}$, results in 
\begin{equation}
	E_l=\frac{\hbar^2}{2m_e r^2} \left(l-\frac{\Phi_B}{\Phi_0^N}\right)^2 \ .
	\label{AB_normal_fluxq}
\end{equation}
Thus for a given energy level, periodic fluctuations occur with increasing magnetic field strength as states with increasing angular momentum are traversed.
Spin can be implemented by doubling the internal degrees of freedom, leading to the Pauli equation with Zeeman interaction and spin-orbit coupling, which are also obtained by relativistic corrections to the Schrödinger equation in the non-relativistic limit of the Dirac equation \cite{SOIcorr}. In SI-units, the Pauli Hamiltonian with relativistic corrections becomes \cite{Berche_2012}, 
\begin{align}
\begin{split}
	H_P &=H_t - \frac{e\hbar}{2m} \boldsymbol{\sigma} \cdot \mathbf{B} - \frac{e\hbar}{4m^2c^2} \boldsymbol{\sigma} \cdot \mathbf{E} \times (\mathbf{p} - e \mathbf{A}) \\&- \frac{e\hbar^2}{8m^2c^2}\left(i\boldsymbol{\sigma} \cdot \bm{\nabla} \times \mathbf{E} + \bm{\nabla} \cdot \mathbf{E} \right),
	\end{split}
	\label{PauliHam}
\end{align}
where the second term is the Zeeman interaction, the third and fourth terms account for spin-orbit coupling, the fifth term is the Darwin term, which can be seen as a kind of spin-orbit term for s-orbitals with zero angular momentum  \cite{Hestenes_consistency}.

\begin{figure}[h!]
	\centering
	\includegraphics[width=0.4\textwidth]{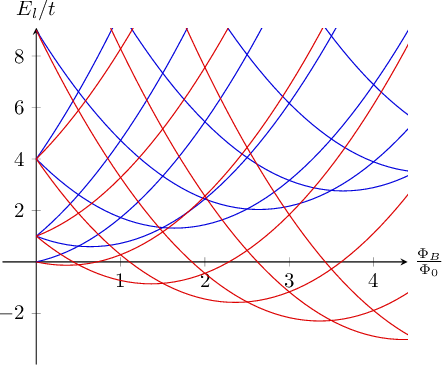}
	\caption{Flux-periodic oscillations of the energy spectrum $E_l$ with $t=\hbar^2/2m_e r^2$ for spin up (red) and spin down (blue) for $\gamma=-0.361$ corresponding to InSb. }
	\label{ab_osc_s}
\end{figure}

Due to the Zeeman interaction the energy bands are shifted to higher and lower energies for spin up and down, Fig. \ref{ab_osc_s}. The wavefunction becomes a composite entity for spin up and spin down, and the Hamiltonian in Eq.\ \eqref{basicHam} with the Zeeman interaction is given by
\begin{equation}
	H_Z =H_t \sigma_0 - |\vec{\mu}| |\vec{B}| \sigma_z,
\end{equation}
where $\sigma_0$ denotes a two by two identity matrix, $\sigma_z$ is a Pauli matrix and $|\vec{\mu}|=\hbar g_ee/4m$ is the magnetic moment for a spin-1/2 particle, given respectively in terms of the (effective) g-factor, charge and mass. For electrons, the spin g-factor has a value\cite{Odom_06} close to 2, however, for carriers in particular materials such as InSb, it becomes effectively much larger due to the contribution from orbital motion \cite{Nilsson2009}.
The energy spectrum becomes spin dependent
\begin{equation}
	E_{\uparrow\downarrow}= E_l \pm \frac{1}{2}g_e\mu_B |\vec{B}|,
	\label{El_updown}
\end{equation}
where $\mu_B=-e\hbar/2m_0$ is the Bohr magneton. 
Defining the energy parameter $t=\hbar^2/2m_e r^2$ and factor $\gamma=m_x g_e/2$, Eq.\eqref{El_updown} can be written as
\begin{equation}
	t^{-1} E_{\uparrow\downarrow}\left(\frac{\Phi_B}{\Phi_0}\right) = \left(l -\frac{\Phi_B}{\Phi_0}^2\right) \pm 2\gamma \frac{\Phi_B}{\Phi_0}.
\end{equation}
In materials, spin-orbit coupling falls into two categories, Rashba\cite{Rashba84} and Dresselhaus\cite{Dresselhaus55}, corresponding to surface and bulk inversion symmetry breaking respectively. In two dimensional systems, the spin-orbit Hamiltonian takes the form
\begin{align}
\begin{split}
		H_{SOI}^{2D} &= \hbar^{-1}[\alpha (\sigma_y p_x - \sigma_x p_y) + \beta (\sigma_x p_x - \sigma_y p_y)] \\
		&= H_{SOI}^R + H_{SOI}^D,
	\end{split}
\end{align}
where $\alpha$ and $\beta$ are the Rashba and Dreselhaus coupling coefficients respectively.
Surface inversion symmetry breaking at a planar interface of two materials results in an uneven charge density, inducing an electric field perpendicular to the interface. The induced field is given by $\mathbf{E}=|\mathbf{E}|\hat{\mathbf{z}}$ if the interface lies in the xy-plane. Then
\begin{equation}
	\frac{\alpha}{\hbar} (\sigma_y p_x - \sigma_x p_y)=	\frac{\alpha}{\hbar}  (\boldsymbol{\sigma} \times \mathbf{p}) \cdot \hat{\mathbf{z}} = -\frac{\alpha}{\hbar E_0} \boldsymbol{\sigma} \cdot (\mathbf{E} \times \mathbf{p}),
	\label{alphaR2deg}
\end{equation}
which shows the similarity to the spin-orbit term in Eq.\ \eqref{PauliHam}.
A core-shell wire with differing core and shell materials can be considered to host two-dimensional electron gas confined to the shell, which is embedded in the radial interface electric field \cite{Bringer2011}. Rashba spin-orbit coupling in such a model becomes an angular form of \eqref{alphaR2deg}, with longitudinal and transverse components for momentum, given in terms of cylindrical coordinates $(r,\phi,z)$,
\begin{equation}
	H_{SOI} = \hbar^{-1} \alpha \left[\sigma_\phi p_z - \sigma_z p_\phi\right] = \hbar^{-1} [H_{SOI}^L + H_{SOI}^T].
	\label{SOItranslong}
\end{equation}

\section{MODEL AND METHODS}
\label{MM}
 \begin{figure}[b]
	\centering
	\includegraphics[width=0.2\textwidth]{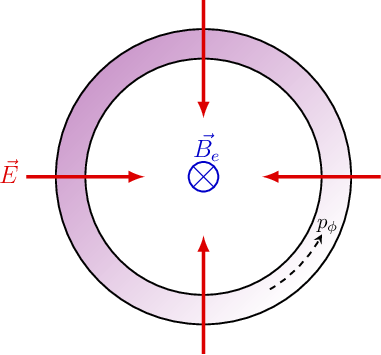}
	\caption{Nanowire cross-section showing the radial interface electric field $\vec{E}$ and applied external magnetic field $\vec{B}$ in the $z$-direction. Angular momentum $p_\phi$ of electron carriers within the shell.}
	\label{NW}
\end{figure}
A semiconductor nanowire shell with proximity-induced superconductivity, situated in an external magnetic field directed parallel to the wire, Fig.\ \ref{NW}, is modelled using the Bogoliubov-de Gennes (BdG) Hamiltonian\cite{Jianxin, Bogoliubov:1958km},
\begin{equation}
	H_{BdG}= 
	\begin{pmatrix}
		[H_w - \mu]&\pm\Delta_{\uparrow\downarrow}\\
		\mp \Delta_{\uparrow\downarrow}^*&-[H^*_{w}-\mu]
	\end{pmatrix},
	\label{BdG}
\end{equation}
written in the particle-hole symmetric Nambu space\cite{Nambu_1960} where $H_w$ is the Hamiltonian of a tubular wire and $\mu$ denotes the chemical potential. 
 The superconducting gap parameter $\pm\Delta_{\uparrow\downarrow}$, couples electrons and holes with opposite spin, effectively describing Cooper pairs in the system. The gap parameter is assumed to be brought on by the proximity effect in the weak coupling limit \cite{TudorSarma}, such that the surrounding superconductor inducing the pairing in the semiconductor shell is not included in the Hamiltonian, in which case the gap parameter is constant, and no phase dependency is imposed on it. At first sight, this may seem as ignoring the Little-Parks oscillations \cite{LittleParks}. However, the phase dependency is attributed to a supercurrent in the surrounding superconductor, described by an effective wavefunction for the whole condensate, Eq.\ \ref{JsGL2}, and neither are included in the current model which focuses on the proximitized semiconductor shell. In the case of a thin superconducting shell (smaller than the coherence length) giving rise to the proximitization, the external magnetic field can penetrate the semiconductor shell as well, leading to $h/2e$ periodicity in magnetoconductance oscillations \cite{GulShapers_2014}. In the current model, the superconducting gap parameter 
 is set as a constant to explore if the BdG spectrum can exhibit $h/e$ or $h/2e$ flux periodicity in this case.
 
 The Hamiltonian for the wire in cylindrical coordinates is given by the sum of transverse, longitudinal, Zeeman and spin-orbit terms respectively,
 \begin{align}
 \begin{split}
 H_w&= H_t + H_l + H_Z + H_{SOI} \\
 &= \frac{(p_{\phi}+eA_{\phi})^2}{2m_e} 
 -\frac{\hbar^2}{2m_er} \frac{\partial}{\partial r} 
 \left(r\frac{\partial}{\partial r}\right)\  \\
 &+ \frac{p_z^2}{2m_e} - g_e \mu_B \sigma B \\
 &+\frac{\alpha}{\hbar}\left[ \sigma_{\phi} p_z -\sigma_z (p_{\phi}+eA_{\phi})\right],
 \end{split}
  \label{Ham_SOI2}
 \end{align}
 where $\sigma=\pm1$ is the spin quantum number.
The symmetric gauge vector potential $A_{\phi}=\frac{1}{2}Br$ incorporates the axial external magnetic field $\mathbf{B}$, giving rise to the Zeeman term $H_Z$ where $g_e$ is the effective Land\'{e} g-factor and $\mu_B$ the Bohr magneton.
The spin-orbit term describes the angular Rashba interaction, arising from a radial electric field induced by crystal asymmetry at the core-shell interface \cite{Bringer2011}. The magnetic vector potential couples linearly to the angular momentum. 

The eigenstates of the wire Hamiltonian \eqref{Ham_SOI2} are written in the basis
\begin{equation}
|g\rangle =|a n \sigma \rangle,
\label{gw}
\end{equation}
where $|a n \sigma \rangle$ denote respectively transverse modes, longitudinal modes and spin. Both finite and infinite wires are modeled by considering longitudinal modes in a sine and exponential basis. The former leading to discrete spectra whilst the latter gives the dispersion relation. For the infinite case, the Hamiltonian becomes a function of the wavevector $\vec{k}$. The eigenstates \eqref{gw} along with the particle-hole space $|\eta\rangle=\pm1$ form a basis $|q\rangle= |\eta g \rangle$ for the BdG Hamiltonian \eqref{BdG} with the following matrix elements. For $\eta=\eta'$
\begin{align}
	\begin{split}
		&\langle q|H_\mathrm{BdG}|q' \rangle=
		\eta[\text{Re}\langle an \sigma |H_\mathrm{w}|a'n' \sigma' \rangle \\&+ i\eta \langle an \sigma|H_\mathrm{w}|a'n' \sigma'\rangle - \mu \delta_{(an \sigma) (a'n' \sigma')}],
	\end{split}
	\label{Diag}
\end{align}
and for $\eta \neq \eta'$,
\begin{equation}
\langle q|H_\mathrm{BdG}|q' \rangle= \eta \sigma \delta_{\sigma,-\sigma'} \delta_{aa'} \delta_{nn'} \Delta_s \ .
\end{equation}
Energy spectra and eigenstates are obtained by numerical diagonalization of a finite-difference approximation using Fortran.
 Parameters for the numerical simulation are chosen to correspond to a proximitized $n$-doped indium antimonide (InSb)\cite{van_Weperen_2015,Gul_2017} semiconducting shell with
 $g_e=-51.6$ and $m_x=0.014$ so that $\gamma=\frac{1}{2} g_e m_x =0.393$. The Rashba spin-orbit interaction parameter is $\alpha=1 \text{ meV nm}$ and the superconducting pairing potential $\Delta= 0.50 \text{ meV}$. The core is considered fully insulating such that the system is effectively a hollow tubular conductor. Analogously to the core/shell system fabricated by Zellekens \textit{et al.}\cite{Zellekens_2020}, hexagonal cross-section geometry is considered. To quantify the effects of cross-section geometry, the hexagonal shell is contrasted with the cases of a cylindrical and square shell. Polygonal shell geometry is implemented by constraints on the polar grid \cite{Sitek15}. The chemical potential $\mu$ is situated within the conduction band. The nanowire radius is $R=50 \text{ nm}$, shell thickness $d=10 \text{ nm}$ and $L_z= 10 \text{ µm}$.
To explore flux-periodic oscillations, the state of the system is calculated for a range of magnetic field strength corresponding to a few flux quanta. 

\section{Cylindrical nanowire shell}
\label{CYL}
\begin{figure} [h!]
	\centering
	\includegraphics[width=0.49\textwidth]{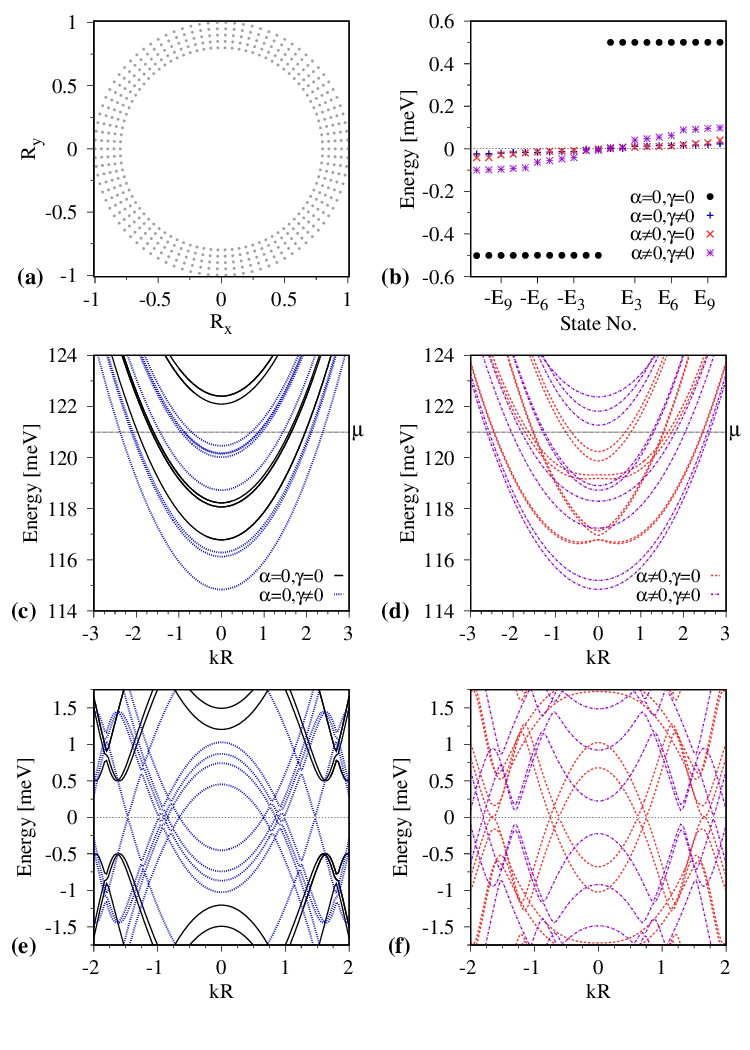}
	\caption{State of the nanowire system at magnetic field strength $|\mathbf{B}|=1.30 \text{ T}$. \textbf{(a)} Circular nanowire cross-section with units $R=50 \text{ nm}$. Shell thickness of $10 \text{ nm}$. \textbf{(b)} Finite wire BdG quasiparticle energy spectra.  \textbf{(c,d)} Energy dispersion of the infinite wire nearby the chemical potential ($\mu$), for the four cases of vanishing/non-vanishing Zeeman interaction ($\gamma$) and spin-orbit interaction ($\alpha$). The dimensionless unit $kR$ denotes the wavevector $k$ times the radius $R$. \textbf{(e,f)} Corresponding BdG quasiparticle energy dispersion for the infinite wire.}
	\label{mm_cyl}
\end{figure}

The eigenstates are initialized at zero magnetic field strength and calculated step-wise up to a field strength of $|\mathbf{B}|= 1.30 \text{ T}$ with an interval of $0.013 \text{ T}$. Basis modes for the wire Hamiltonian are 9 transverse and 200 longitudinal. The chemical potential is situated above three spin-degenerate states in the absence of Zeeman and spin-orbit interaction. Four cases of vanishing/non-vanishing Zeeman and spin-orbit interaction are compared at the maximum magnetic field strength, to highlight spin-dependent effects.
The grid cross-section is shown in Fig.\ \ref{mm_cyl}(a) and the associated finite wire BdG energy spectra in Fig.\ \ref{mm_cyl}(b). Separately including either the Zeeman or the spin-orbit interaction, leads to similar spectra of the lowest energy states. Including both interactions leads to an increase in the separation between non-degenerate eigenvalues, due to the combined spin-splitting effect. 
The Zeeman interaction lifts the spin-degeneracy and as a results the aligned spin becomes the lowest energy state in the spectrum of the infinite wire, Fig.\ \ref{mm_cyl}(c). The spin-orbit interaction involves two terms, Eq.\ \eqref{Ham_SOI2}, the former of which can be considered to give an effective orbital magnetic field whilst the latter leads to an effective axial magnetic field, contributing to the Zeeman splitting. Hence, both a horizontal shift and a vertical split in the two lowest energy opposite spin states is obtained, including the spin-orbit interaction on its own, Fig.\ \ref{mm_cyl}(c). 
Combining the Zeeman and Rashba spin-orbit interaction, the Zeeman splitting dominates. As one component of the magnetic field associated to the spin-orbit interaction is parallel to the external magnetic field, Eq.\ \eqref{Ham_SOI2}, additional spin-splitting occurs for each flux quantum. The closing of the BdG gap in the infinite wire dispersion varies considerably between the cases, Figs.\ \ref{mm_cyl}(e,f). In the case of a vanishing Zeeman and spin-orbit interaction, the gap never closes over the magnetic field cycle. In all other cases the gap closes, but at different momentum values. 

\begin{figure}[h!]
	\centering
	\includegraphics[width=0.49\textwidth]{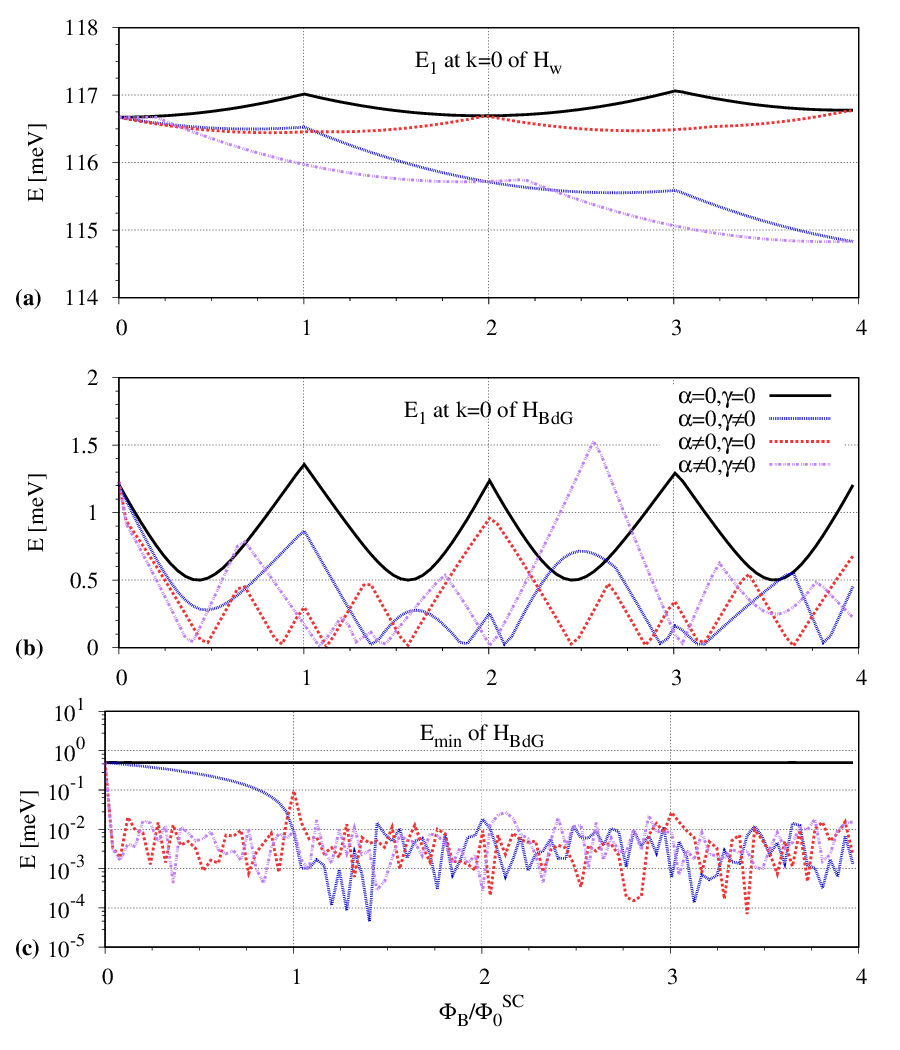}
	\caption{Cylindrical wire: Flux-periodic oscillations of the lowest energy state, in terms of the flux quantum $\Phi_0^{SC}=h/2e$, for the four cases of vanishing/non-vanishing Zeeman $\gamma$, and spin-orbit interaction $\alpha$. \textbf{(a)} The normal conducting wire, both finite and infinite at $k=0$. \textbf{(b)} Superconducting infinite wire at $k=0$. \textbf{(c)}  Superconducting infinite wire for all $k$.}
	\label{LP_cyl}
\end{figure}
Fig.\ \ref{LP_cyl} shows the flux-periodic oscillations of the lowest energy state of the finite and infinite normal- and superconducting wire at $k=0$.  The oscillations of the normal conducting wire, Fig.\ \ref{LP_cyl}(a), are the same for the finite and infinite wire and display flux-periodicity of $2\Phi_0^{SC}=h/e$, in accordance with Eq.\ \ref{AB_normal_fluxq}. The Rashba spin-orbit coupling has the effect of shifting the oscillations in phase whilst the Zeeman interaction introduces a linear dependence of the flux quantum in addition to the quadratic dependence, without altering the periodicity. Combing the two interactions merges these effects in an expected manner.  The lowest energy state of the infinite wire BdG Hamiltonian at $k=0$, Fig.\ \ref{LP_cyl}(b), displays clear $h/2e$-periodic oscillations in the absence of both Zeeman and spin-orbit interaction. Adding the Zeeman interaction mixes bands of opposite spins, resulting in sub-oscillations with alternating inverted curvature profile, beyond a magnetic field strength corresponding to one superconducting flux quantum. The spin-orbit interaction unevenly doubles the periodicity and sharpens the oscillations of all periods, in accordance with horizontal splitting of the dispersion, Fig.\ \ref{mm_cyl}(b). The combination of the Zeeman and spin-orbit interaction results in approximately alternating constructive and destructive coherence of the separate oscillations, breaking the periodicity. The detailed shape of the oscillations depends on the position of the chemical potential, which governs the transition between states of varying angular momentum.

Fig.\ \ref{LP_cyl}(c) shows the minimum energy for all values of the wavevector $k$, which is in direct correspondence with the lowest energy state of the finite wire. In the absence of both Zeeman and spin-orbit interaction, the minimum gap is independent of the magnetic field strength, but rapidly closes if the spin-dependent interactions are included. The closing is linear in the case of the Zeeman interaction and happens at $k\ne0$, which is also the case for the spin-orbit interaction but then the closing is abrupt. Incidentally, a small gap reopening and closing occurs around one and three flux quanta, in relation to the smallest amplitudes of the flux-periodic oscillations in Fig.\ \ref{LP_cyl}(b). Irregularities of the minimum energy in Fig.\ \ref{LP_cyl}(c) are due to small variations of the gap for differing values of the wavevector $k$.
\section{Effects of polygonal cross-section geometry}
\label{POLY}
\begin{figure} [h!]
	\centering
	\includegraphics[width=0.49\textwidth]{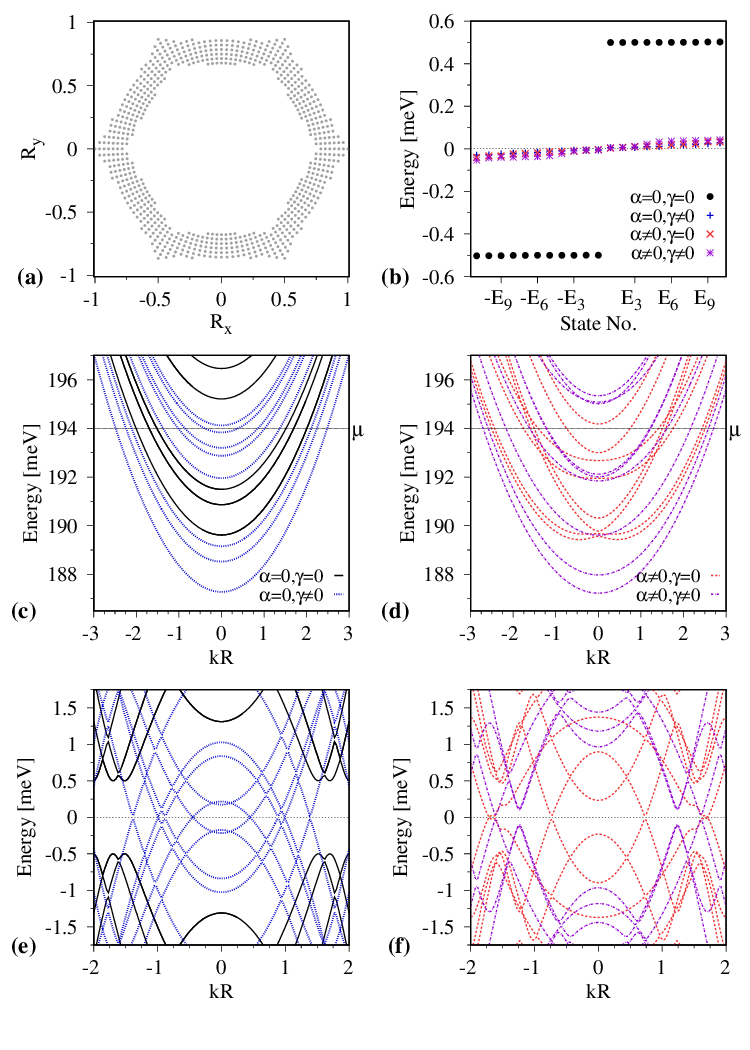}
	\caption{State of the nanowire system at magnetic field strength $|\mathbf{B}|=1.56 \text{ T}$. \textbf{(a)} Hexagonal nanowire cross-section with units $R=50 \text{ nm}$. Minimal shell thickness of $10 \text{ nm}$. \textbf{(b)} Finite wire BdG energy spectra. \textbf{(c,d)} Energy dispersion of the infinite wire nearby the chemical potential ($\mu$), for the four cases of vanishing/non-vanishing Zeeman interaction ($\gamma$) and spin-orbit interaction ($\alpha$). The dimensionless unit $kR$ denotes the wavevector times the radius. \textbf{(e,f)} Corresponding BdG quasiparticle energy dispersion for the infinite wire.}
	\label{mm_hex}
\end{figure}

We proceed by comparing with a hexagonal shell, and for a fair comparison, the minimal shell thickness is set equal to the cylindrical shell. In this way, the sides are of similar thickness to the cylindrical case, whilst the corner radius is slightly larger, Fig.\ \ref{mm_hex}(a). The cylindrical coordinate system of the grid results in irregularities of the shell surface, which is practical with respect to the irregularities of fabricated samples,  as they are rarely perfectly regular. In the finite wire BdG spectrum, Fig.\ \ref{mm_hex}(b), a minor decrease of variance between the cases of vanishing/nonvanishing Zeeman and spin-orbit interaction can be seen compared with the cylindrical wire. However, the systems in Fig.\ \ref{mm_cyl} and Fig.\ \ref{mm_hex} are at slightly different values of magnetic field strength $|\vec{B}|$, as higher magnetic fields are needed for the same number of flux quanta due to the smaller area, $A$, of the hexagon, as seen from the definition of magnetic flux $\Phi_B=|\vec{B}|A$.
By introducing corners into the grid domain of the transverse Hamiltonian, separation of the lowest energy levels is induced, Figs.\ \ref{mm_hex}(c,d), due to the corner localization \cite{Sitek15}, which increases the difference in wave vectors $\vec{k}=\hbar^{-1} \vec{p}$, for adjacent energy levels at the given chemical potential. This results in an increased difference in momentum between the quasiparticle bands, Figs.\ \ref{mm_hex}(e,f), compared with the cylindrical case.  

\begin{figure}[h!]
	\centering
	\includegraphics[width=0.49\textwidth]{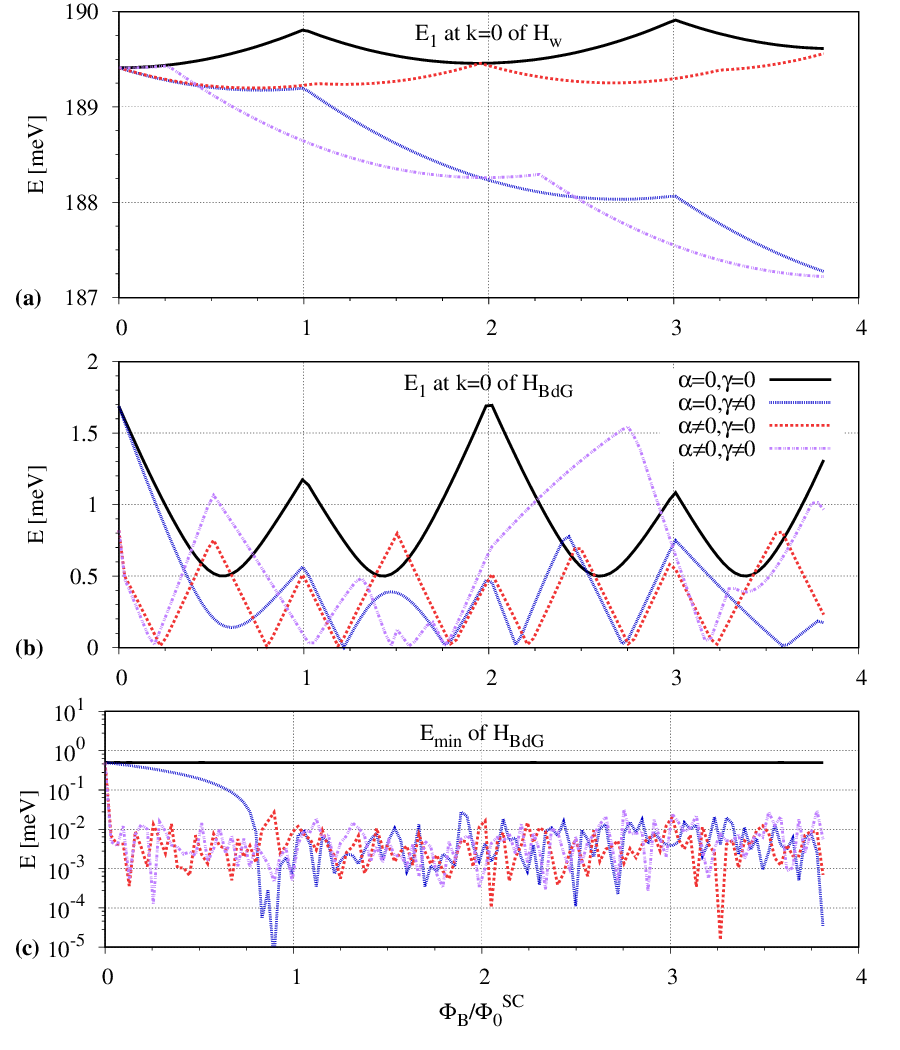}
	\caption{Hexagonal wire: Flux-periodic oscillations of the lowest energy state for a hexagonal wire cross-section, in terms of the flux quantum $\Phi_0^{SC}=h/2e$, for the four cases of vanishing/non-vanishing Zeeman $\gamma$, and spin-orbit interaction $\alpha$. \textbf{(a)} The normal conducting wire, both finite and infinite at $k=0$. \textbf{(b)} Superconducting infinite wire at $k=0$. \textbf{(c)}  Superconducting infinite wire for all $k$, direct correspondence with the lowest energy state of the finite wire.}
	\label{LP_hex}
\end{figure}

Compared with the cylindrical case, the flux oscillations of the normal conducting wire are nearly identical, Fig.\ \ref{LP_hex}(a). This is expected as the effective quantum wells formed by the corners, overlap in the sides \cite{Sitek15,Ferrari09b,Wu_92}, so that the hexagon is not too different from the cylinder. More specifically, the angular momentum from the external magnetic field is not overpowered by the corner localization. Therefore the theory for cylindrical nanowires may often be in qualitative agreement with experimental results from hexagonal wires \cite{SanJose_2022}.
Note, however, that the angular momentum is not conserved for the hexagonal case so the oscillations are not strictly periodic \cite{Ballester_2013,Torres_2018}.

\begin{figure}[h!]
	\centering
	\includegraphics[width=0.49\textwidth]{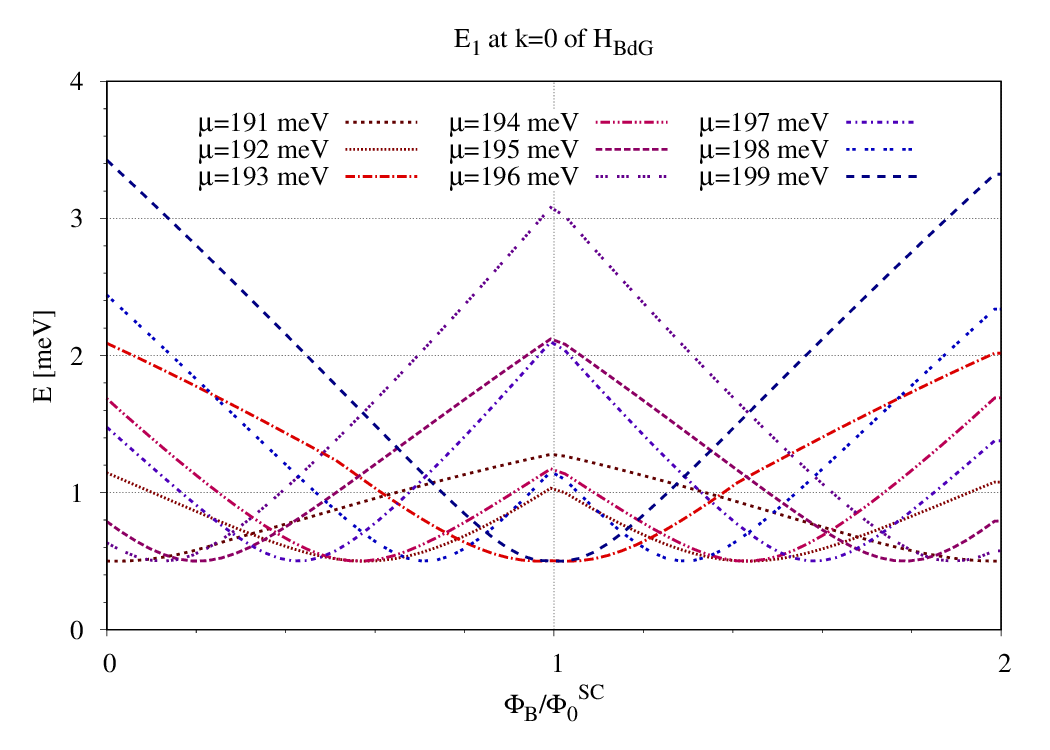}
	\caption{Varying period of flux-periodic oscillations of the lowest energy state of the infinite wire BdG spectra at $k=0$ of the hexagonal nanowire.}
	\label{chempot_FQosc}
\end{figure}

\begin{figure}[h!]
	\centering
	\includegraphics[width=0.49\textwidth]{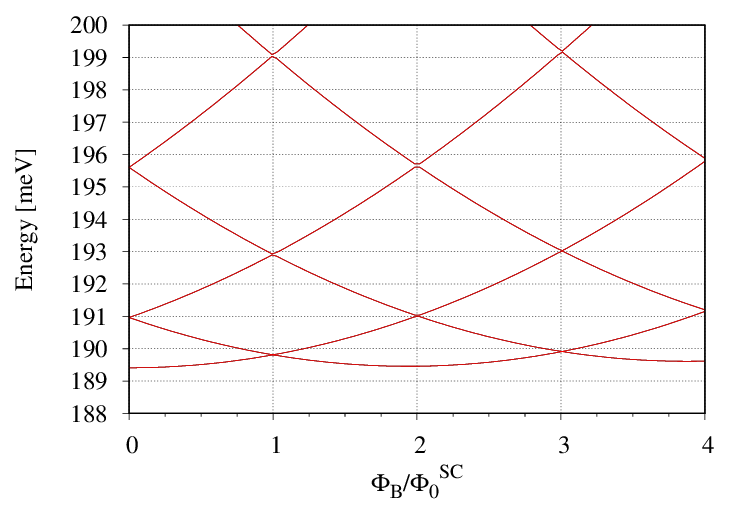}
	\caption{Oscillations in the energy spectrum of the infinite normal conducting hexagonal wire.}
	\label{HexOscPhi}
\end{figure}

\begin{figure}[h!]
	\centering
	\includegraphics[width=0.5\textwidth]{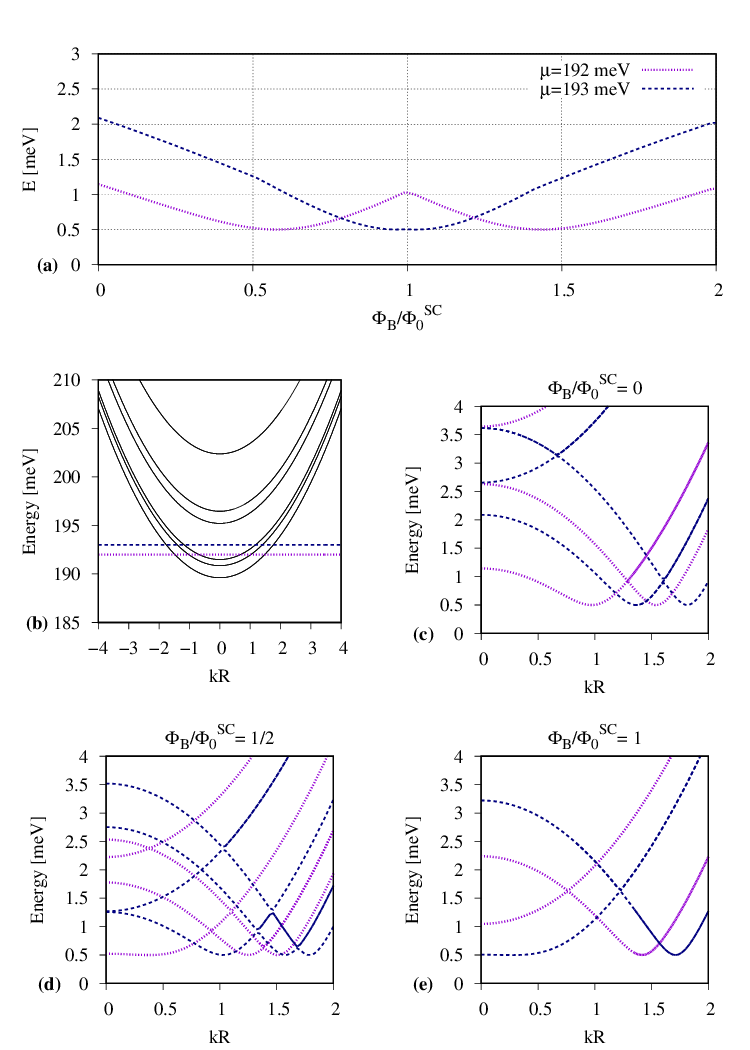}
	\caption{(a) BdG gap at $k=0$ of the infinite wire for two values of the chemical potential. (b) Infinite wire dispersion. (c) BdG dispersion for magnetic fields strengths equal to zero superconducting flux quanta, (d) half a flux quantum, (e) one flux quantum.}
	\label{4p1_hex_FP}
\end{figure}

The oscillations of the BdG spectrum of the infinite wire at $k=0$, Fig.\ \ref{LP_hex}(b,c), share the same features as the cylindrical case.  A period of $h/2e$ is clearly obtained for vanishing Zeeman and spin-orbit interaction but the peak amplitudes are alternately equal.
The $h/2e$ periodicity is found to be highly dependent on the chemical potential. This applies as well to the cylindrical case, but is explored here in more detail for the hexagonal cross-section, as it is a more realistic and common configuration of nanowire geometry. 
\FloatBarrier
In  Fig.\ \ref{chempot_FQosc}, the periodicity in the case of vanishing Zeeman and spin-orbit interaction is shown for nine values of the chemical potential in integer steps. With increasing energy, the periodicity approximately alternates between periodicity of a normal and superconducting flux quantum.
Comparing with the flux-periodic oscillation of the normal conducting energy bands, Fig.\ \ref{HexOscPhi}, there is a correspondence with values of chemical potentials at crossings and $h/e$ periodicity seen in Fig.\ \ref{chempot_FQosc}. This is clearly seen at $\mu= 199 \text{ meV}$ for example, where the period minima corresponds to the crossing at one flux quantum in Fig.\ \ref{HexOscPhi}. Conversely, for $\mu=191 \text{ meV}$, the crossing is at zero and three flux quanta, leading to a period maxima at one flux quantum. As the chemical potential is shifted away from crossings, the sub-periods appear. At values exactly midway between crossings, where $\mu=192 \text{ meV}$ comes close, the $h/2e$ periodicity is displayed.
To specify the underlying interaction in the BdG spectra, we chose two chemical potential values and compare the corresponding flux-periodicity, Fig.\ \ref{4p1_hex_FP}(a). 
The BdG Hamiltonian essentially doubles the degrees of freedom, coupling electrons and holes of opposite spin by the gap parameter. The value of the chemical potential in the normal wire, Fig.\ \ref{4p1_hex_FP}(b) determines around which values of $k$ the induced gap is situated. The key difference between the two cases is that in the case of the higher chemical potential, there is a crossing of the adjacent upper energy level in the BdG spectra, Fig.\ \ref{4p1_hex_FP}(c), such that the curvature of the second band at $k=0$ is inverted with respect to the first one. As the magnetic field strength is increased to the value of half a flux quantum, this energy level meets the lowest one at the energy value $1.25 \text{ meV}$, Fig.\ \ref{4p1_hex_FP}(d). Further increasing the magnetic field strength, the two bands join and the lowest energy is overtaken in a sense so that the energy keeps lowering up to the value of one flux quantum, Fig.\ \ref{4p1_hex_FP}(e). A similar process has been reported in the flux periodicity of quantum rings with spin degrees of freedom included, giving rise to the fractional Aharanov-Bohm effect\cite{Niemela_1996}.

\begin{figure} [h!]
		\centering
	\includegraphics[width=0.49\textwidth]{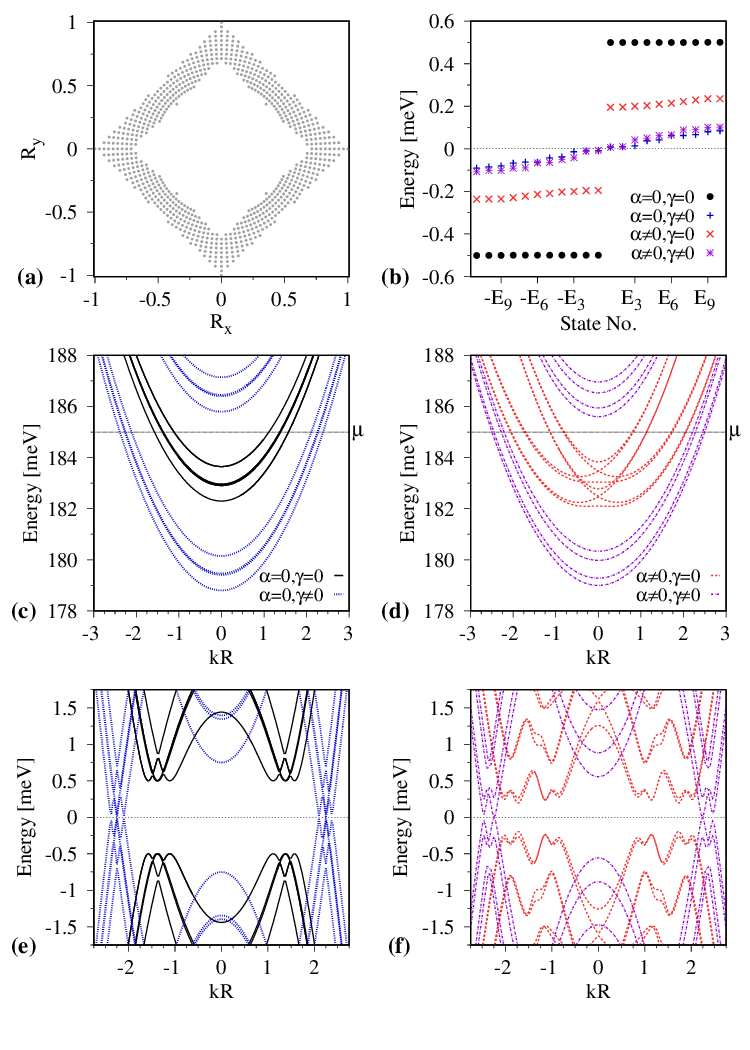}
	\caption{State of the nanowire system at magnetic field strength $|\mathbf{B}|=2.32 \text{ T}$. \textbf{(a)} Square nanowire cross-section with units $R=50 \text{ nm}$. Minimal shell thickness of $10 \text{ nm}$. \textbf{(b)} Finite wire BdG quasiparticle energy spectra. \textbf{(c,d)} Energy dispersion of the infinite wire nearby the chemical potential ($\mu$), for the four cases of vanishing/non-vanishing Zeeman interaction ($\gamma$) and spin-orbit interaction ($\alpha$). The dimensionless unit $kR$ denotes the wavevector $k \text{ [m]}^{-1}$ times radius $R \text{ [m]}$. \textbf{(e,f)} Corresponding BdG quasiparticle energy dispersion for the infinite wire.}
	\label{mm_square}
\end{figure}

To investigate the effects of increased corner localization, the energy values for a square cross-section are calculated, Figs.\ \ref{mm_square}(a,b). As before, higher magnetic field strength is needed for an equal amount of flux quanta, due to the decreased area of the square cross-section. The corner states form a near degenerate group that is still influenced by the orbital magnetic term, which manifests in the splitting of the lowest energy states in the absence of both $\gamma, \alpha$, Fig.\ \ref{mm_square}(c). The corner geometry has the non-trivial effect \cite{Bringer2011} of enhancing the longitudinal component of the spin-orbit coupling, Eq.\ \eqref{SOItranslong}, leading to crossings in more energy levels, Fig.\ \ref{mm_square}(d). Analogously to the foregoing cases, including the Zeeman interaction allows for gap closing of the BdG dispersion, Fig.\ \ref{mm_square}(e), as the magnetic field overpowers the proximity-induced coupling of opposite spin states. In the case of solely nonvanishing spin-orbit coupling there is no gap closing at the given magnetic field strength, Fig.\ \ref{mm_square}(f), since the corner localization decreases the transverse momentum \cite{Andrei}.

\begin{figure}[h!]
	\centering
	\includegraphics[width=0.49\textwidth]{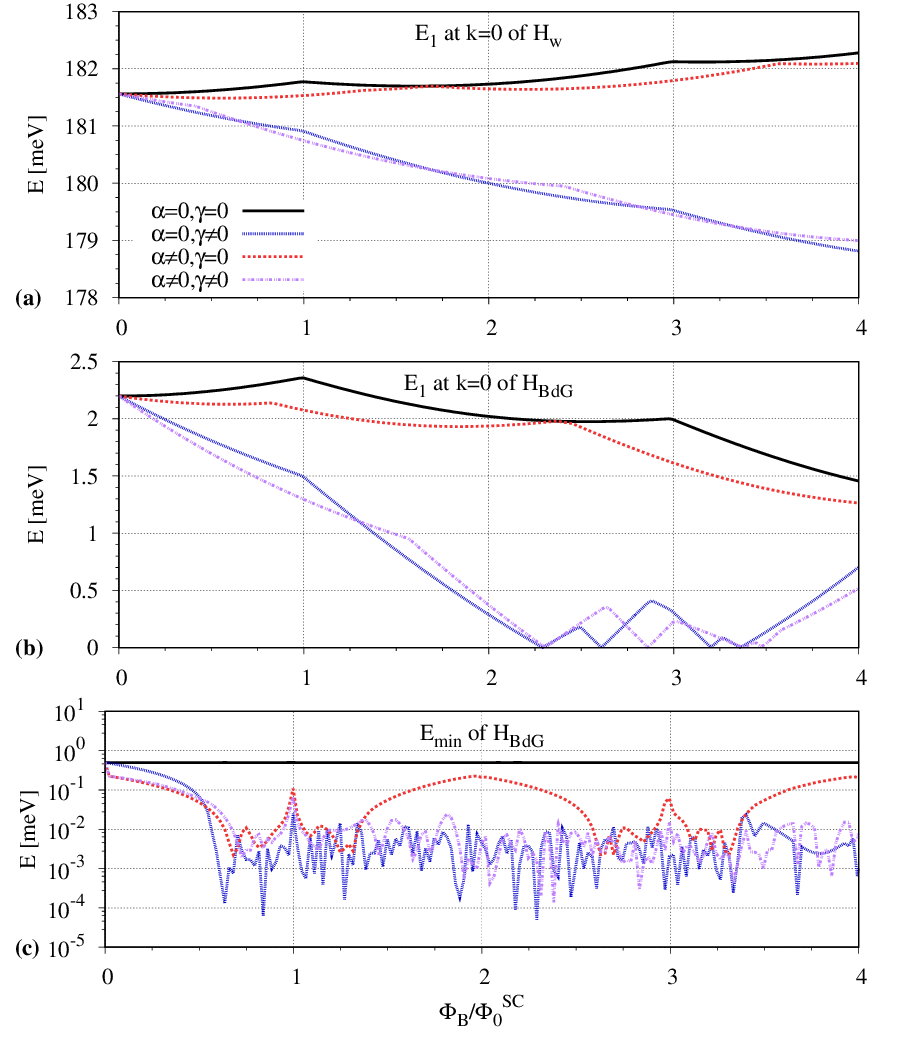}
	\caption{Flux-periodic oscillations of the lowest energy state for a square wire cross-section, in terms of the flux quantum $\Phi_0^{SC}=h/2e$, for the four cases of vanishing/non-vanishing Zeeman $\gamma$, and spin-orbit interaction $\alpha$. \textbf{(a)} The normal conducting wire, both finite and infinite at $k=0$. \textbf{(b)} Superconducting infinite wire at $k=0$. \textbf{(c)}  Superconducting infinite wire for all $k$.}
	\label{LP_square}
\end{figure}

The flux-periodic oscillations of the normal conducting wire are nearly flattened out due to corner localization Fig.\ \ref{LP_square}(a). Furthermore, there is very little indication of the superconducting flux quantum in Fig.\ \ref{LP_square}(b). This is attributed to the large separation between the first groups of corner- and side-localized states, as this is the main variable between the cases. 
The oscillations of the minimum energy for all $k$ in the infinite wire BdG spectra, Fig.\ \ref{LP_square}(c) show alternating gap opening and closing with periodicity of the superconducting flux quantum. The induced magnetic field from the spin-orbit coupling opposes the external magnetic field, allowing for the gap opening.
\FloatBarrier

\section{CONCLUSIONS}
\label{CONCL}
Flux-periodic oscillations in a core-shell nanowire system with proximity induced superconductivity are explored numerically. 
We explore the instance where supercurrents are absent in the semiconductor shell, as in the case of full-shell covering of the nanowire by a thick superconductor.
Magneto-oscillations of the lowest energy states of the Bogoliubov-de Gennes Hamiltonian (BdG) and its components are calculated for cylindrical, hexagonal and square cross-section geometries. For the former two cases, $h/2e$ periodicity is found for specific values of the chemical potential whilst the periodicity for the square wire is found to be strictly $h/e$.
The superconducting gap parameter is constant throughout the calculation, with no phase dependency imposed. Zeeman and spin-orbit interaction are found to influence the magneto-oscillations considerably so that  a periodicity of a fractional superconducting flux quantum is obtained at a zero wavevector in the infinite wire BdG dispersion.
The flux-periodicity arises from minimal coupling in the diagonal components of the BdG Hamiltonian.  The transition between $h/e$ and $h/2e$ periodicity is found to be dependent on the adjacent higher energy level with respect to the chemical potential. 

In contrast to the case where supercurrents impose $h/2e$ oscillations, corresponding to a $2e$ charge unit, we have shown that in their absence, the magneto-oscillations are determined by the spectrum of the semiconductor nanowire in the normal state. If the nanowire has a minimal deviation from cylindrical symmetry, the oscillations can have the periodicity of $h/e$ or $h/2e$, depending on the position of the chemical potential relative to the energy bands.  If the cylindrical symmetry is sufficiently broken, solely $h/e$ periodicity is obtained. The results show in what way the chemical potential, which is controllable by a gate voltage, may influence flux-periodicity in hybrid semiconductor-superconductor devices.
\\

\begin{acknowledgments}
This research was supported by the Reykjavik University Research Fund, project no.\ 218043 and the Icelandic Research Fund, grant no. 206568-051. We are grateful to Patrick Zellekens and Vidar Gudmundsson for discussions. 
\end{acknowledgments}
\bibliographystyle{apsrev4-1}
\bibliography{CSNreferences_LP_SOI}

\begin{thebibliography}{71}%
\makeatletter
\providecommand \@ifxundefined [1]{%
 \@ifx{#1\undefined}
}%
\providecommand \@ifnum [1]{%
 \ifnum #1\expandafter \@firstoftwo
 \else \expandafter \@secondoftwo
 \fi
}%
\providecommand \@ifx [1]{%
 \ifx #1\expandafter \@firstoftwo
 \else \expandafter \@secondoftwo
 \fi
}%
\providecommand \natexlab [1]{#1}%
\providecommand \enquote  [1]{``#1''}%
\providecommand \bibnamefont  [1]{#1}%
\providecommand \bibfnamefont [1]{#1}%
\providecommand \citenamefont [1]{#1}%
\providecommand \href@noop [0]{\@secondoftwo}%
\providecommand \href [0]{\begingroup \@sanitize@url \@href}%
\providecommand \@href[1]{\@@startlink{#1}\@@href}%
\providecommand \@@href[1]{\endgroup#1\@@endlink}%
\providecommand \@sanitize@url [0]{\catcode `\\12\catcode `\$12\catcode
  `\&12\catcode `\#12\catcode `\^12\catcode `\_12\catcode `\%12\relax}%
\providecommand \@@startlink[1]{}%
\providecommand \@@endlink[0]{}%
\providecommand \url  [0]{\begingroup\@sanitize@url \@url }%
\providecommand \@url [1]{\endgroup\@href {#1}{\urlprefix }}%
\providecommand \urlprefix  [0]{URL }%
\providecommand \Eprint [0]{\href }%
\providecommand \doibase [0]{http://dx.doi.org/}%
\providecommand \selectlanguage [0]{\@gobble}%
\providecommand \bibinfo  [0]{\@secondoftwo}%
\providecommand \bibfield  [0]{\@secondoftwo}%
\providecommand \translation [1]{[#1]}%
\providecommand \BibitemOpen [0]{}%
\providecommand \bibitemStop [0]{}%
\providecommand \bibitemNoStop [0]{.\EOS\space}%
\providecommand \EOS [0]{\spacefactor3000\relax}%
\providecommand \BibitemShut  [1]{\csname bibitem#1\endcsname}%
\let\auto@bib@innerbib\@empty
\bibitem [{\citenamefont {Zhu}\ and\ \citenamefont {Wang}(1995)}]{Zhu_95}%
  \BibitemOpen
  \bibfield  {author} {\bibinfo {author} {\bibfnamefont {J.-X.}\ \bibnamefont
  {Zhu}}\ and\ \bibinfo {author} {\bibfnamefont {Z.~D.}\ \bibnamefont {Wang}},\
  }\href {\doibase 10.1103/PhysRevB.51.13813} {\bibfield  {journal} {\bibinfo
  {journal} {Phys. Rev. B}\ }\textbf {\bibinfo {volume} {51}},\ \bibinfo
  {pages} {13813} (\bibinfo {year} {1995})}\BibitemShut {NoStop}%
\bibitem [{\citenamefont {Byers}\ and\ \citenamefont {Yang}(1961)}]{Yang_61}%
  \BibitemOpen
  \bibfield  {author} {\bibinfo {author} {\bibfnamefont {N.}~\bibnamefont
  {Byers}}\ and\ \bibinfo {author} {\bibfnamefont {C.~N.}\ \bibnamefont
  {Yang}},\ }\href {\doibase 10.1103/PhysRevLett.7.46} {\bibfield  {journal}
  {\bibinfo  {journal} {Phys. Rev. Lett.}\ }\textbf {\bibinfo {volume} {7}},\
  \bibinfo {pages} {46} (\bibinfo {year} {1961})}\BibitemShut {NoStop}%
\bibitem [{\citenamefont {Little}\ and\ \citenamefont
  {Parks}(1962)}]{LittleParks}%
  \BibitemOpen
  \bibfield  {author} {\bibinfo {author} {\bibfnamefont {W.~A.}\ \bibnamefont
  {Little}}\ and\ \bibinfo {author} {\bibfnamefont {R.~D.}\ \bibnamefont
  {Parks}},\ }\href {\doibase 10.1103/PhysRevLett.9.9} {\bibfield  {journal}
  {\bibinfo  {journal} {Phys. Rev. Lett.}\ }\textbf {\bibinfo {volume} {9}},\
  \bibinfo {pages} {9–12} (\bibinfo {year} {1962})}\BibitemShut {NoStop}%
\bibitem [{\citenamefont {Malozovsky}\ and\ \citenamefont
  {Fan}(1999)}]{Fan_1999}%
  \BibitemOpen
  \bibfield  {author} {\bibinfo {author} {\bibfnamefont {Y.}~\bibnamefont
  {Malozovsky}}\ and\ \bibinfo {author} {\bibfnamefont {J.}~\bibnamefont
  {Fan}},\ }\href {\doibase https://doi.org/10.1016/S0375-9601(99)00250-9}
  {\bibfield  {journal} {\bibinfo  {journal} {Physics Letters A}\ }\textbf
  {\bibinfo {volume} {257}},\ \bibinfo {pages} {332} (\bibinfo {year}
  {1999})}\BibitemShut {NoStop}%
\bibitem [{\citenamefont {{Peñaranda}}\ \emph {et~al.}(2019)\citenamefont
  {{Peñaranda}}, \citenamefont {{Aguado}}, \citenamefont {{San-Jose}},\ and\
  \citenamefont {{Prada}}}]{evenodd}%
  \BibitemOpen
  \bibfield  {author} {\bibinfo {author} {\bibfnamefont {F.}~\bibnamefont
  {{Peñaranda}}}, \bibinfo {author} {\bibfnamefont {R.}~\bibnamefont
  {{Aguado}}}, \bibinfo {author} {\bibfnamefont {P.}~\bibnamefont
  {{San-Jose}}}, \ and\ \bibinfo {author} {\bibfnamefont {E.}~\bibnamefont
  {{Prada}}},\ }\href@noop {} {\bibfield  {journal} {\bibinfo  {journal} {arXiv
  e-prints}\ ,\ \bibinfo {eid} {arXiv:1911.06805}} (\bibinfo {year} {2019})},\
  \Eprint {http://arxiv.org/abs/1911.06805} {arXiv:1911.06805
  [cond-mat.mes-hall]} \BibitemShut {NoStop}%
\bibitem [{\citenamefont {Hajer}\ \emph {et~al.}(2019)\citenamefont {Hajer},
  \citenamefont {Kessel}, \citenamefont {Brüne}, \citenamefont {Stehno},
  \citenamefont {Buhmann},\ and\ \citenamefont
  {Molenkamp}}]{2019ProxCoreShell}%
  \BibitemOpen
  \bibfield  {author} {\bibinfo {author} {\bibfnamefont {J.}~\bibnamefont
  {Hajer}}, \bibinfo {author} {\bibfnamefont {M.}~\bibnamefont {Kessel}},
  \bibinfo {author} {\bibfnamefont {C.}~\bibnamefont {Brüne}}, \bibinfo
  {author} {\bibfnamefont {M.}~\bibnamefont {Stehno}}, \bibinfo {author}
  {\bibfnamefont {H.}~\bibnamefont {Buhmann}}, \ and\ \bibinfo {author}
  {\bibfnamefont {L.}~\bibnamefont {Molenkamp}},\ }\href {\doibase
  10.1021/acs.nanolett.9b01472} {\bibfield  {journal} {\bibinfo  {journal}
  {Nano Letters}\ } (\bibinfo {year} {2019}),\
  10.1021/acs.nanolett.9b01472}\BibitemShut {NoStop}%
\bibitem [{\citenamefont {Bl{\"o}mers}\ \emph {et~al.}(2013)\citenamefont
  {Bl{\"o}mers}, \citenamefont {Rieger}, \citenamefont {Zellekens},
  \citenamefont {Haas}, \citenamefont {Lepsa}, \citenamefont {Hardtdegen},
  \citenamefont {G{\"u}l}, \citenamefont {Demarina}, \citenamefont
  {Gr{\"u}tzmacher}, \citenamefont {L{\"u}th},\ and\ \citenamefont
  {Sch{\"a}pers}}]{Blomers_13}%
  \BibitemOpen
  \bibfield  {author} {\bibinfo {author} {\bibfnamefont {C.}~\bibnamefont
  {Bl{\"o}mers}}, \bibinfo {author} {\bibfnamefont {T.}~\bibnamefont {Rieger}},
  \bibinfo {author} {\bibfnamefont {P.}~\bibnamefont {Zellekens}}, \bibinfo
  {author} {\bibfnamefont {F.}~\bibnamefont {Haas}}, \bibinfo {author}
  {\bibfnamefont {M.~I.}\ \bibnamefont {Lepsa}}, \bibinfo {author}
  {\bibfnamefont {H.}~\bibnamefont {Hardtdegen}}, \bibinfo {author}
  {\bibfnamefont {{\"O}.}~\bibnamefont {G{\"u}l}}, \bibinfo {author}
  {\bibfnamefont {N.}~\bibnamefont {Demarina}}, \bibinfo {author}
  {\bibfnamefont {D.}~\bibnamefont {Gr{\"u}tzmacher}}, \bibinfo {author}
  {\bibfnamefont {H.}~\bibnamefont {L{\"u}th}}, \ and\ \bibinfo {author}
  {\bibfnamefont {T.}~\bibnamefont {Sch{\"a}pers}},\ }\href@noop {} {\bibfield
  {journal} {\bibinfo  {journal} {Nanotechnology}\ }\textbf {\bibinfo {volume}
  {24}},\ \bibinfo {pages} {035203} (\bibinfo {year} {2013})}\BibitemShut
  {NoStop}%
\bibitem [{\citenamefont {Pistol}\ and\ \citenamefont
  {Pryor}(2008)}]{Pistol_2008}%
  \BibitemOpen
  \bibfield  {author} {\bibinfo {author} {\bibfnamefont {M.-E.}\ \bibnamefont
  {Pistol}}\ and\ \bibinfo {author} {\bibfnamefont {C.~E.}\ \bibnamefont
  {Pryor}},\ }\href {\doibase 10.1103/PhysRevB.78.115319} {\bibfield  {journal}
  {\bibinfo  {journal} {Phys. Rev. B}\ }\textbf {\bibinfo {volume} {78}},\
  \bibinfo {pages} {115319} (\bibinfo {year} {2008})}\BibitemShut {NoStop}%
\bibitem [{\citenamefont {Rieger}\ \emph {et~al.}(2012)\citenamefont {Rieger},
  \citenamefont {Luysberg}, \citenamefont {Schäpers}, \citenamefont
  {Grützmacher},\ and\ \citenamefont {Lepsa}}]{Rieger2012}%
  \BibitemOpen
  \bibfield  {author} {\bibinfo {author} {\bibfnamefont {T.}~\bibnamefont
  {Rieger}}, \bibinfo {author} {\bibfnamefont {M.}~\bibnamefont {Luysberg}},
  \bibinfo {author} {\bibfnamefont {T.}~\bibnamefont {Schäpers}}, \bibinfo
  {author} {\bibfnamefont {D.}~\bibnamefont {Grützmacher}}, \ and\ \bibinfo
  {author} {\bibfnamefont {M.~I.}\ \bibnamefont {Lepsa}},\ }\href {\doibase
  10.1021/nl302502b} {\bibfield  {journal} {\bibinfo  {journal} {Nano Letters}\
  }\textbf {\bibinfo {volume} {12}},\ \bibinfo {pages} {5559–5564} (\bibinfo
  {year} {2012})}\BibitemShut {NoStop}%
\bibitem [{\citenamefont {Fan}\ \emph {et~al.}(2006)\citenamefont {Fan},
  \citenamefont {Knez}, \citenamefont {Scholz}, \citenamefont {Nielsch},
  \citenamefont {Pippel}, \citenamefont {Hesse}, \citenamefont {Gösele},\ and\
  \citenamefont {Zacharias}}]{Fan_06}%
  \BibitemOpen
  \bibfield  {author} {\bibinfo {author} {\bibfnamefont {H.}~\bibnamefont
  {Fan}}, \bibinfo {author} {\bibfnamefont {M.}~\bibnamefont {Knez}}, \bibinfo
  {author} {\bibfnamefont {R.}~\bibnamefont {Scholz}}, \bibinfo {author}
  {\bibfnamefont {K.}~\bibnamefont {Nielsch}}, \bibinfo {author} {\bibfnamefont
  {E.}~\bibnamefont {Pippel}}, \bibinfo {author} {\bibfnamefont
  {D.}~\bibnamefont {Hesse}}, \bibinfo {author} {\bibfnamefont
  {U.}~\bibnamefont {Gösele}}, \ and\ \bibinfo {author} {\bibfnamefont
  {M.}~\bibnamefont {Zacharias}},\ }\href
  {http://stacks.iop.org/0957-4484/17/i=20/a=020} {\bibfield  {journal}
  {\bibinfo  {journal} {Nanotechnology}\ }\textbf {\bibinfo {volume} {17}},\
  \bibinfo {pages} {5157} (\bibinfo {year} {2006})}\BibitemShut {NoStop}%
\bibitem [{\citenamefont {{Göransson}}\ \emph {et~al.}(2019)\citenamefont
  {{Göransson}}, \citenamefont {{Heurlin}}, \citenamefont {{Dalelkhan}},
  \citenamefont {{Abay}}, \citenamefont {{Messing}}, \citenamefont {{Maisi}},
  \citenamefont {{Borgström}},\ and\ \citenamefont {{Xu}}}]{TriHex}%
  \BibitemOpen
  \bibfield  {author} {\bibinfo {author} {\bibfnamefont {D.~J.~O.}\
  \bibnamefont {{Göransson}}}, \bibinfo {author} {\bibfnamefont
  {M.}~\bibnamefont {{Heurlin}}}, \bibinfo {author} {\bibfnamefont
  {B.}~\bibnamefont {{Dalelkhan}}}, \bibinfo {author} {\bibfnamefont
  {S.}~\bibnamefont {{Abay}}}, \bibinfo {author} {\bibfnamefont {M.~E.}\
  \bibnamefont {{Messing}}}, \bibinfo {author} {\bibfnamefont {V.~F.}\
  \bibnamefont {{Maisi}}}, \bibinfo {author} {\bibfnamefont {M.~T.}\
  \bibnamefont {{Borgström}}}, \ and\ \bibinfo {author} {\bibfnamefont
  {H.~Q.}\ \bibnamefont {{Xu}}},\ }\href {\doibase 10.1063/1.5084222}
  {\bibfield  {journal} {\bibinfo  {journal} {Applied Physics Letters}\
  }\textbf {\bibinfo {volume} {114}},\ \bibinfo {eid} {053108} (\bibinfo {year}
  {2019})}\BibitemShut {NoStop}%
\bibitem [{\citenamefont {Ferrari}\ \emph {et~al.}(2009)\citenamefont
  {Ferrari}, \citenamefont {Goldoni}, \citenamefont {Bertoni}, \citenamefont
  {Cuoghi},\ and\ \citenamefont {Molinari}}]{Ferrari09b}%
  \BibitemOpen
  \bibfield  {author} {\bibinfo {author} {\bibfnamefont {G.}~\bibnamefont
  {Ferrari}}, \bibinfo {author} {\bibfnamefont {G.}~\bibnamefont {Goldoni}},
  \bibinfo {author} {\bibfnamefont {A.}~\bibnamefont {Bertoni}}, \bibinfo
  {author} {\bibfnamefont {G.}~\bibnamefont {Cuoghi}}, \ and\ \bibinfo {author}
  {\bibfnamefont {E.}~\bibnamefont {Molinari}},\ }\href@noop {} {\bibfield
  {journal} {\bibinfo  {journal} {Nano Letters}\ }\textbf {\bibinfo {volume}
  {9}},\ \bibinfo {pages} {1631} (\bibinfo {year} {2009})}\BibitemShut
  {NoStop}%
\bibitem [{\citenamefont {Sitek}\ \emph {et~al.}(2015)\citenamefont {Sitek},
  \citenamefont {Serra}, \citenamefont {Gudmundsson},\ and\ \citenamefont
  {Manolescu}}]{Sitek15}%
  \BibitemOpen
  \bibfield  {author} {\bibinfo {author} {\bibfnamefont {A.}~\bibnamefont
  {Sitek}}, \bibinfo {author} {\bibfnamefont {L.}~\bibnamefont {Serra}},
  \bibinfo {author} {\bibfnamefont {V.}~\bibnamefont {Gudmundsson}}, \ and\
  \bibinfo {author} {\bibfnamefont {A.}~\bibnamefont {Manolescu}},\ }\href@noop
  {} {\bibfield  {journal} {\bibinfo  {journal} {Phys. Rev. B}\ }\textbf
  {\bibinfo {volume} {91}},\ \bibinfo {pages} {235429} (\bibinfo {year}
  {2015})}\BibitemShut {NoStop}%
\bibitem [{\citenamefont {Manolescu}\ \emph {et~al.}(2017)\citenamefont
  {Manolescu}, \citenamefont {Sitek}, \citenamefont {Osca}, \citenamefont
  {Serra}, \citenamefont {Gudmundsson},\ and\ \citenamefont
  {Stanescu}}]{Andrei}%
  \BibitemOpen
  \bibfield  {author} {\bibinfo {author} {\bibfnamefont {A.}~\bibnamefont
  {Manolescu}}, \bibinfo {author} {\bibfnamefont {A.}~\bibnamefont {Sitek}},
  \bibinfo {author} {\bibfnamefont {J.}~\bibnamefont {Osca}}, \bibinfo {author}
  {\bibfnamefont {L.}~\bibnamefont {Serra}}, \bibinfo {author} {\bibfnamefont
  {V.}~\bibnamefont {Gudmundsson}}, \ and\ \bibinfo {author} {\bibfnamefont
  {T.~D.}\ \bibnamefont {Stanescu}},\ }\href@noop {} {\bibfield  {journal}
  {\bibinfo  {journal} {Phys. Rev. B}\ }\textbf {\bibinfo {volume} {96}},\
  \bibinfo {pages} {125435} (\bibinfo {year} {2017})}\BibitemShut {NoStop}%
\bibitem [{\citenamefont {Zellekens}\ \emph {et~al.}(2020)\citenamefont
  {Zellekens}, \citenamefont {Demarina}, \citenamefont {Jan{\ss}en},
  \citenamefont {Rieger}, \citenamefont {Lepsa}, \citenamefont {Perla},
  \citenamefont {Panaitov}, \citenamefont {Lüth}, \citenamefont
  {Grützmacher},\ and\ \citenamefont {Schäpers}}]{Zellekens_2020}%
  \BibitemOpen
  \bibfield  {author} {\bibinfo {author} {\bibfnamefont {P.}~\bibnamefont
  {Zellekens}}, \bibinfo {author} {\bibfnamefont {N.}~\bibnamefont {Demarina}},
  \bibinfo {author} {\bibfnamefont {J.}~\bibnamefont {Jan{\ss}en}}, \bibinfo
  {author} {\bibfnamefont {T.}~\bibnamefont {Rieger}}, \bibinfo {author}
  {\bibfnamefont {M.~I.}\ \bibnamefont {Lepsa}}, \bibinfo {author}
  {\bibfnamefont {P.}~\bibnamefont {Perla}}, \bibinfo {author} {\bibfnamefont
  {G.}~\bibnamefont {Panaitov}}, \bibinfo {author} {\bibfnamefont
  {H.}~\bibnamefont {Lüth}}, \bibinfo {author} {\bibfnamefont
  {D.}~\bibnamefont {Grützmacher}}, \ and\ \bibinfo {author} {\bibfnamefont
  {T.}~\bibnamefont {Schäpers}},\ }\href {\doibase 10.1088/1361-6641/ab8396}
  {\bibfield  {journal} {\bibinfo  {journal} {Semiconductor Science and
  Technology}\ }\textbf {\bibinfo {volume} {35}},\ \bibinfo {pages} {085003}
  (\bibinfo {year} {2020})}\BibitemShut {NoStop}%
\bibitem [{\citenamefont {Önder Gül}\ \emph {et~al.}(2015)\citenamefont
  {Önder Gül}, \citenamefont {van Woerkom}, \citenamefont {van Weperen},
  \citenamefont {Car}, \citenamefont {Plissard}, \citenamefont {Bakkers},\ and\
  \citenamefont {Kouwenhoven}}]{G_l_2015}%
  \BibitemOpen
  \bibfield  {author} {\bibinfo {author} {\bibnamefont {Önder Gül}}, \bibinfo
  {author} {\bibfnamefont {D.~J.}\ \bibnamefont {van Woerkom}}, \bibinfo
  {author} {\bibfnamefont {I.}~\bibnamefont {van Weperen}}, \bibinfo {author}
  {\bibfnamefont {D.}~\bibnamefont {Car}}, \bibinfo {author} {\bibfnamefont
  {S.~R.}\ \bibnamefont {Plissard}}, \bibinfo {author} {\bibfnamefont {E.~P.
  A.~M.}\ \bibnamefont {Bakkers}}, \ and\ \bibinfo {author} {\bibfnamefont
  {L.~P.}\ \bibnamefont {Kouwenhoven}},\ }\href {\doibase
  10.1088/0957-4484/26/21/215202} {\bibfield  {journal} {\bibinfo  {journal}
  {Nanotechnology}\ }\textbf {\bibinfo {volume} {26}},\ \bibinfo {pages}
  {215202} (\bibinfo {year} {2015})}\BibitemShut {NoStop}%
\bibitem [{\citenamefont {Wójcik}\ \emph {et~al.}(2019)\citenamefont
  {Wójcik}, \citenamefont {Bertoni},\ and\ \citenamefont
  {Goldoni}}]{Wojcik2019}%
  \BibitemOpen
  \bibfield  {author} {\bibinfo {author} {\bibfnamefont {P.}~\bibnamefont
  {Wójcik}}, \bibinfo {author} {\bibfnamefont {A.}~\bibnamefont {Bertoni}}, \
  and\ \bibinfo {author} {\bibfnamefont {G.}~\bibnamefont {Goldoni}},\ }\href
  {\doibase 10.1063/1.5082602} {\bibfield  {journal} {\bibinfo  {journal}
  {Applied Physics Letters}\ }\textbf {\bibinfo {volume} {114}},\ \bibinfo
  {pages} {073102} (\bibinfo {year} {2019})}\BibitemShut {NoStop}%
\bibitem [{\citenamefont {van Weperen}\ \emph {et~al.}(2015)\citenamefont {van
  Weperen}, \citenamefont {Tarasinski}, \citenamefont {Eeltink}, \citenamefont
  {Pribiag}, \citenamefont {Plissard}, \citenamefont {Bakkers}, \citenamefont
  {Kouwenhoven},\ and\ \citenamefont {Wimmer}}]{van_Weperen_2015}%
  \BibitemOpen
  \bibfield  {author} {\bibinfo {author} {\bibfnamefont {I.}~\bibnamefont {van
  Weperen}}, \bibinfo {author} {\bibfnamefont {B.}~\bibnamefont {Tarasinski}},
  \bibinfo {author} {\bibfnamefont {D.}~\bibnamefont {Eeltink}}, \bibinfo
  {author} {\bibfnamefont {V.~S.}\ \bibnamefont {Pribiag}}, \bibinfo {author}
  {\bibfnamefont {S.~R.}\ \bibnamefont {Plissard}}, \bibinfo {author}
  {\bibfnamefont {E.~P. A.~M.}\ \bibnamefont {Bakkers}}, \bibinfo {author}
  {\bibfnamefont {L.~P.}\ \bibnamefont {Kouwenhoven}}, \ and\ \bibinfo {author}
  {\bibfnamefont {M.}~\bibnamefont {Wimmer}},\ }\href {\doibase
  10.1103/physrevb.91.201413} {\bibfield  {journal} {\bibinfo  {journal}
  {Physical Review B}\ }\textbf {\bibinfo {volume} {91}} (\bibinfo {year}
  {2015}),\ 10.1103/physrevb.91.201413}\BibitemShut {NoStop}%
\bibitem [{\citenamefont {Alicea}(2012)}]{Alicea_2012}%
  \BibitemOpen
  \bibfield  {author} {\bibinfo {author} {\bibfnamefont {J.}~\bibnamefont
  {Alicea}},\ }\href@noop {} {\bibfield  {journal} {\bibinfo  {journal}
  {Reports on Progress in Physics}\ }\textbf {\bibinfo {volume} {75}},\
  \bibinfo {pages} {076501} (\bibinfo {year} {2012})}\BibitemShut {NoStop}%
\bibitem [{\citenamefont {Stanescu}\ \emph {et~al.}(2011)\citenamefont
  {Stanescu}, \citenamefont {Lutchyn},\ and\ \citenamefont {{Das
  Sarma}}}]{Stanescu2011}%
  \BibitemOpen
  \bibfield  {author} {\bibinfo {author} {\bibfnamefont {T.~D.}\ \bibnamefont
  {Stanescu}}, \bibinfo {author} {\bibfnamefont {R.~M.}\ \bibnamefont
  {Lutchyn}}, \ and\ \bibinfo {author} {\bibfnamefont {S.}~\bibnamefont {{Das
  Sarma}}},\ }\href {\doibase 10.1103/PhysRevB.84.144522} {\bibfield  {journal}
  {\bibinfo  {journal} {Phys. Rev. B}\ }\textbf {\bibinfo {volume} {84}},\
  \bibinfo {pages} {144522} (\bibinfo {year} {2011})}\BibitemShut {NoStop}%
\bibitem [{\citenamefont {Zhang}\ \emph {et~al.}(2019)\citenamefont {Zhang},
  \citenamefont {Liu}, \citenamefont {Wimmer},\ and\ \citenamefont
  {Kouwenhoven}}]{Zhang2019}%
  \BibitemOpen
  \bibfield  {author} {\bibinfo {author} {\bibfnamefont {H.}~\bibnamefont
  {Zhang}}, \bibinfo {author} {\bibfnamefont {D.~E.}\ \bibnamefont {Liu}},
  \bibinfo {author} {\bibfnamefont {M.}~\bibnamefont {Wimmer}}, \ and\ \bibinfo
  {author} {\bibfnamefont {L.~P.}\ \bibnamefont {Kouwenhoven}},\ }\href
  {\doibase 10.1038/s41467-019-13133-1} {\bibfield  {journal} {\bibinfo
  {journal} {Nature Communications}\ }\textbf {\bibinfo {volume} {10}},\
  \bibinfo {pages} {5128} (\bibinfo {year} {2019})}\BibitemShut {NoStop}%
\bibitem [{\citenamefont {Tsuei}\ \emph {et~al.}(1996)\citenamefont {Tsuei},
  \citenamefont {Kirtley}, \citenamefont {Gupta}, \citenamefont {Sun},
  \citenamefont {Moler}, \citenamefont {Ren},\ and\ \citenamefont
  {Wang}}]{Tsuei_1996}%
  \BibitemOpen
  \bibfield  {author} {\bibinfo {author} {\bibfnamefont {C.~C.}\ \bibnamefont
  {Tsuei}}, \bibinfo {author} {\bibfnamefont {J.~R.}\ \bibnamefont {Kirtley}},
  \bibinfo {author} {\bibfnamefont {A.}~\bibnamefont {Gupta}}, \bibinfo
  {author} {\bibfnamefont {J.~Z.}\ \bibnamefont {Sun}}, \bibinfo {author}
  {\bibfnamefont {K.~A.}\ \bibnamefont {Moler}}, \bibinfo {author}
  {\bibfnamefont {Z.~F.}\ \bibnamefont {Ren}}, \ and\ \bibinfo {author}
  {\bibfnamefont {J.~H.}\ \bibnamefont {Wang}},\ }\href {\doibase
  10.1088/0031-8949/1996/t66/038} {\bibfield  {journal} {\bibinfo  {journal}
  {Physica Scripta}\ }\textbf {\bibinfo {volume} {T66}},\ \bibinfo {pages}
  {212} (\bibinfo {year} {1996})}\BibitemShut {NoStop}%
\bibitem [{\citenamefont {Fourcade}(1986)}]{Fourcade_86}%
  \BibitemOpen
  \bibfield  {author} {\bibinfo {author} {\bibfnamefont {B.}~\bibnamefont
  {Fourcade}},\ }\href {\doibase 10.1103/PhysRevB.33.6644} {\bibfield
  {journal} {\bibinfo  {journal} {Phys. Rev. B}\ }\textbf {\bibinfo {volume}
  {33}},\ \bibinfo {pages} {6644} (\bibinfo {year} {1986})}\BibitemShut
  {NoStop}%
\bibitem [{\citenamefont {Li}\ and\ \citenamefont
  {Soukoulis}(1986)}]{Qiming_86}%
  \BibitemOpen
  \bibfield  {author} {\bibinfo {author} {\bibfnamefont {Q.}~\bibnamefont
  {Li}}\ and\ \bibinfo {author} {\bibfnamefont {C.~M.}\ \bibnamefont
  {Soukoulis}},\ }\href {\doibase 10.1103/PhysRevLett.57.3105} {\bibfield
  {journal} {\bibinfo  {journal} {Phys. Rev. Lett.}\ }\textbf {\bibinfo
  {volume} {57}},\ \bibinfo {pages} {3105} (\bibinfo {year}
  {1986})}\BibitemShut {NoStop}%
\bibitem [{\citenamefont {Pannetier}\ \emph {et~al.}(1985)\citenamefont
  {Pannetier}, \citenamefont {Chaussy}, \citenamefont {Rammal},\ and\
  \citenamefont {Gandit}}]{Pannatier_85}%
  \BibitemOpen
  \bibfield  {author} {\bibinfo {author} {\bibfnamefont {B.}~\bibnamefont
  {Pannetier}}, \bibinfo {author} {\bibfnamefont {J.}~\bibnamefont {Chaussy}},
  \bibinfo {author} {\bibfnamefont {R.}~\bibnamefont {Rammal}}, \ and\ \bibinfo
  {author} {\bibfnamefont {P.}~\bibnamefont {Gandit}},\ }\href {\doibase
  10.1103/PhysRevB.31.3209} {\bibfield  {journal} {\bibinfo  {journal} {Phys.
  Rev. B}\ }\textbf {\bibinfo {volume} {31}},\ \bibinfo {pages} {3209}
  (\bibinfo {year} {1985})}\BibitemShut {NoStop}%
\bibitem [{\citenamefont {Altshuler}\ \emph {et~al.}(1981)\citenamefont
  {Altshuler}, \citenamefont {Aronov},\ and\ \citenamefont {Spivak}}]{AAS_81}%
  \BibitemOpen
  \bibfield  {author} {\bibinfo {author} {\bibfnamefont {B.}~\bibnamefont
  {Altshuler}}, \bibinfo {author} {\bibfnamefont {A.}~\bibnamefont {Aronov}}, \
  and\ \bibinfo {author} {\bibfnamefont {B.}~\bibnamefont {Spivak}},\
  }\href@noop {} {\bibfield  {journal} {\bibinfo  {journal} {JETP Lett.}\
  }\textbf {\bibinfo {volume} {33}},\ \bibinfo {pages} {94} (\bibinfo {year}
  {1981})}\BibitemShut {NoStop}%
\bibitem [{\citenamefont {Seidel}\ and\ \citenamefont {Lee}(2005)}]{Seidel_05}%
  \BibitemOpen
  \bibfield  {author} {\bibinfo {author} {\bibfnamefont {A.}~\bibnamefont
  {Seidel}}\ and\ \bibinfo {author} {\bibfnamefont {D.-H.}\ \bibnamefont
  {Lee}},\ }\href {\doibase 10.1103/PhysRevB.71.045113} {\bibfield  {journal}
  {\bibinfo  {journal} {Phys. Rev. B}\ }\textbf {\bibinfo {volume} {71}},\
  \bibinfo {pages} {045113} (\bibinfo {year} {2005})}\BibitemShut {NoStop}%
\bibitem [{\citenamefont {Zha}\ \emph {et~al.}(2012)\citenamefont {Zha},
  \citenamefont {Wang}, \citenamefont {Wang},\ and\ \citenamefont
  {Zhou}}]{Zha_2012}%
  \BibitemOpen
  \bibfield  {author} {\bibinfo {author} {\bibfnamefont {G.-Q.}\ \bibnamefont
  {Zha}}, \bibinfo {author} {\bibfnamefont {S.-S.}\ \bibnamefont {Wang}},
  \bibinfo {author} {\bibfnamefont {J.-C.}\ \bibnamefont {Wang}}, \ and\
  \bibinfo {author} {\bibfnamefont {S.-P.}\ \bibnamefont {Zhou}},\ }\href
  {\doibase 10.1063/1.4742051} {\bibfield  {journal} {\bibinfo  {journal}
  {Journal of Applied Physics}\ }\textbf {\bibinfo {volume} {112}},\ \bibinfo
  {pages} {033907} (\bibinfo {year} {2012})}\BibitemShut {NoStop}%
\bibitem [{\citenamefont {Loder}\ \emph {et~al.}(2008)\citenamefont {Loder},
  \citenamefont {Kampf}, \citenamefont {Kopp}, \citenamefont {Mannhart},
  \citenamefont {Schneider},\ and\ \citenamefont {Barash}}]{Loder2008}%
  \BibitemOpen
  \bibfield  {author} {\bibinfo {author} {\bibfnamefont {F.}~\bibnamefont
  {Loder}}, \bibinfo {author} {\bibfnamefont {A.~P.}\ \bibnamefont {Kampf}},
  \bibinfo {author} {\bibfnamefont {T.}~\bibnamefont {Kopp}}, \bibinfo {author}
  {\bibfnamefont {J.}~\bibnamefont {Mannhart}}, \bibinfo {author}
  {\bibfnamefont {C.~W.}\ \bibnamefont {Schneider}}, \ and\ \bibinfo {author}
  {\bibfnamefont {Y.~S.}\ \bibnamefont {Barash}},\ }\href {\doibase
  10.1038/nphys813} {\bibfield  {journal} {\bibinfo  {journal} {Nature
  Physics}\ }\textbf {\bibinfo {volume} {4}},\ \bibinfo {pages} {112} (\bibinfo
  {year} {2008})}\BibitemShut {NoStop}%
\bibitem [{\citenamefont {Xu}\ \emph {et~al.}(2020)\citenamefont {Xu},
  \citenamefont {Li},\ and\ \citenamefont {Chien}}]{Xu_2020}%
  \BibitemOpen
  \bibfield  {author} {\bibinfo {author} {\bibfnamefont {X.}~\bibnamefont
  {Xu}}, \bibinfo {author} {\bibfnamefont {Y.}~\bibnamefont {Li}}, \ and\
  \bibinfo {author} {\bibfnamefont {C.~L.}\ \bibnamefont {Chien}},\ }\href
  {\doibase 10.1103/PhysRevLett.124.167001} {\bibfield  {journal} {\bibinfo
  {journal} {Phys. Rev. Lett.}\ }\textbf {\bibinfo {volume} {124}},\ \bibinfo
  {pages} {167001} (\bibinfo {year} {2020})}\BibitemShut {NoStop}%
\bibitem [{\citenamefont {Yasui}\ \emph {et~al.}(2017)\citenamefont {Yasui},
  \citenamefont {Lahabi}, \citenamefont {Anwar}, \citenamefont {Nakamura},
  \citenamefont {Yonezawa}, \citenamefont {Terashima}, \citenamefont {Aarts},\
  and\ \citenamefont {Maeno}}]{Yasui_2017}%
  \BibitemOpen
  \bibfield  {author} {\bibinfo {author} {\bibfnamefont {Y.}~\bibnamefont
  {Yasui}}, \bibinfo {author} {\bibfnamefont {K.}~\bibnamefont {Lahabi}},
  \bibinfo {author} {\bibfnamefont {M.~S.}\ \bibnamefont {Anwar}}, \bibinfo
  {author} {\bibfnamefont {Y.}~\bibnamefont {Nakamura}}, \bibinfo {author}
  {\bibfnamefont {S.}~\bibnamefont {Yonezawa}}, \bibinfo {author}
  {\bibfnamefont {T.}~\bibnamefont {Terashima}}, \bibinfo {author}
  {\bibfnamefont {J.}~\bibnamefont {Aarts}}, \ and\ \bibinfo {author}
  {\bibfnamefont {Y.}~\bibnamefont {Maeno}},\ }\href {\doibase
  10.1103/PhysRevB.96.180507} {\bibfield  {journal} {\bibinfo  {journal} {Phys.
  Rev. B}\ }\textbf {\bibinfo {volume} {96}},\ \bibinfo {pages} {180507}
  (\bibinfo {year} {2017})}\BibitemShut {NoStop}%
\bibitem [{\citenamefont {Cai}\ \emph {et~al.}(2013)\citenamefont {Cai},
  \citenamefont {Ying}, \citenamefont {Staley}, \citenamefont {Xin},
  \citenamefont {Fobes}, \citenamefont {Liu}, \citenamefont {Mao},\ and\
  \citenamefont {Liu}}]{Cai_13}%
  \BibitemOpen
  \bibfield  {author} {\bibinfo {author} {\bibfnamefont {X.}~\bibnamefont
  {Cai}}, \bibinfo {author} {\bibfnamefont {Y.~A.}\ \bibnamefont {Ying}},
  \bibinfo {author} {\bibfnamefont {N.~E.}\ \bibnamefont {Staley}}, \bibinfo
  {author} {\bibfnamefont {Y.}~\bibnamefont {Xin}}, \bibinfo {author}
  {\bibfnamefont {D.}~\bibnamefont {Fobes}}, \bibinfo {author} {\bibfnamefont
  {T.~J.}\ \bibnamefont {Liu}}, \bibinfo {author} {\bibfnamefont {Z.~Q.}\
  \bibnamefont {Mao}}, \ and\ \bibinfo {author} {\bibfnamefont
  {Y.}~\bibnamefont {Liu}},\ }\href {\doibase 10.1103/PhysRevB.87.081104}
  {\bibfield  {journal} {\bibinfo  {journal} {Phys. Rev. B}\ }\textbf {\bibinfo
  {volume} {87}},\ \bibinfo {pages} {081104} (\bibinfo {year}
  {2013})}\BibitemShut {NoStop}%
\bibitem [{\citenamefont {Nazarov}(1989)}]{Nazarov_89}%
  \BibitemOpen
  \bibfield  {author} {\bibinfo {author} {\bibfnamefont {Y.~V.}\ \bibnamefont
  {Nazarov}},\ }\href@noop {} {\bibfield  {journal} {\bibinfo  {journal} {Zh.
  Eksp. Teor. Fiz.}\ }\textbf {\bibinfo {volume} {95}},\ \bibinfo {pages} {338}
  (\bibinfo {year} {1989})}\BibitemShut {NoStop}%
\bibitem [{\citenamefont {Geshkenbein}\ \emph {et~al.}(1987)\citenamefont
  {Geshkenbein}, \citenamefont {Larkin},\ and\ \citenamefont
  {Barone}}]{Geshkenbein_87}%
  \BibitemOpen
  \bibfield  {author} {\bibinfo {author} {\bibfnamefont {V.~B.}\ \bibnamefont
  {Geshkenbein}}, \bibinfo {author} {\bibfnamefont {A.~I.}\ \bibnamefont
  {Larkin}}, \ and\ \bibinfo {author} {\bibfnamefont {A.}~\bibnamefont
  {Barone}},\ }\href {\doibase 10.1103/PhysRevB.36.235} {\bibfield  {journal}
  {\bibinfo  {journal} {Phys. Rev. B}\ }\textbf {\bibinfo {volume} {36}},\
  \bibinfo {pages} {235} (\bibinfo {year} {1987})}\BibitemShut {NoStop}%
\bibitem [{\citenamefont {Ivanov}(2001)}]{Ivanov_2001}%
  \BibitemOpen
  \bibfield  {author} {\bibinfo {author} {\bibfnamefont {D.~A.}\ \bibnamefont
  {Ivanov}},\ }\href {\doibase 10.1103/PhysRevLett.86.268} {\bibfield
  {journal} {\bibinfo  {journal} {Phys. Rev. Lett.}\ }\textbf {\bibinfo
  {volume} {86}},\ \bibinfo {pages} {268} (\bibinfo {year} {2001})}\BibitemShut
  {NoStop}%
\bibitem [{\citenamefont {Seo}\ \emph {et~al.}(2015)\citenamefont {Seo},
  \citenamefont {Kang}, \citenamefont {Kwon},\ and\ \citenamefont
  {Shin}}]{Seo_2015}%
  \BibitemOpen
  \bibfield  {author} {\bibinfo {author} {\bibfnamefont {S.~W.}\ \bibnamefont
  {Seo}}, \bibinfo {author} {\bibfnamefont {S.}~\bibnamefont {Kang}}, \bibinfo
  {author} {\bibfnamefont {W.~J.}\ \bibnamefont {Kwon}}, \ and\ \bibinfo
  {author} {\bibfnamefont {Y.-i.}\ \bibnamefont {Shin}},\ }\href {\doibase
  10.1103/PhysRevLett.115.015301} {\bibfield  {journal} {\bibinfo  {journal}
  {Phys. Rev. Lett.}\ }\textbf {\bibinfo {volume} {115}},\ \bibinfo {pages}
  {015301} (\bibinfo {year} {2015})}\BibitemShut {NoStop}%
\bibitem [{\citenamefont {Manni}\ \emph {et~al.}(2012)\citenamefont {Manni},
  \citenamefont {Lagoudakis}, \citenamefont {Liew}, \citenamefont {Andr{\'e}},
  \citenamefont {Savona},\ and\ \citenamefont {Deveaud}}]{Manni_2012}%
  \BibitemOpen
  \bibfield  {author} {\bibinfo {author} {\bibfnamefont {F.}~\bibnamefont
  {Manni}}, \bibinfo {author} {\bibfnamefont {K.~G.}\ \bibnamefont
  {Lagoudakis}}, \bibinfo {author} {\bibfnamefont {T.~C.~H.}\ \bibnamefont
  {Liew}}, \bibinfo {author} {\bibfnamefont {R.}~\bibnamefont {Andr{\'e}}},
  \bibinfo {author} {\bibfnamefont {V.}~\bibnamefont {Savona}}, \ and\ \bibinfo
  {author} {\bibfnamefont {B.}~\bibnamefont {Deveaud}},\ }\href {\doibase
  10.1038/ncomms2310} {\bibfield  {journal} {\bibinfo  {journal} {Nature
  Communications}\ }\textbf {\bibinfo {volume} {3}},\ \bibinfo {pages} {1309}
  (\bibinfo {year} {2012})}\BibitemShut {NoStop}%
\bibitem [{\citenamefont {Autti}\ \emph {et~al.}(2016)\citenamefont {Autti},
  \citenamefont {Dmitriev}, \citenamefont {M\"akinen}, \citenamefont
  {Soldatov}, \citenamefont {Volovik}, \citenamefont {Yudin}, \citenamefont
  {Zavjalov},\ and\ \citenamefont {Eltsov}}]{Autti_2016}%
  \BibitemOpen
  \bibfield  {author} {\bibinfo {author} {\bibfnamefont {S.}~\bibnamefont
  {Autti}}, \bibinfo {author} {\bibfnamefont {V.~V.}\ \bibnamefont {Dmitriev}},
  \bibinfo {author} {\bibfnamefont {J.~T.}\ \bibnamefont {M\"akinen}}, \bibinfo
  {author} {\bibfnamefont {A.~A.}\ \bibnamefont {Soldatov}}, \bibinfo {author}
  {\bibfnamefont {G.~E.}\ \bibnamefont {Volovik}}, \bibinfo {author}
  {\bibfnamefont {A.~N.}\ \bibnamefont {Yudin}}, \bibinfo {author}
  {\bibfnamefont {V.~V.}\ \bibnamefont {Zavjalov}}, \ and\ \bibinfo {author}
  {\bibfnamefont {V.~B.}\ \bibnamefont {Eltsov}},\ }\href {\doibase
  10.1103/PhysRevLett.117.255301} {\bibfield  {journal} {\bibinfo  {journal}
  {Phys. Rev. Lett.}\ }\textbf {\bibinfo {volume} {117}},\ \bibinfo {pages}
  {255301} (\bibinfo {year} {2016})}\BibitemShut {NoStop}%
\bibitem [{\citenamefont {Salomaa}\ and\ \citenamefont
  {Volovik}(1987)}]{Volovik_1987}%
  \BibitemOpen
  \bibfield  {author} {\bibinfo {author} {\bibfnamefont {M.~M.}\ \bibnamefont
  {Salomaa}}\ and\ \bibinfo {author} {\bibfnamefont {G.~E.}\ \bibnamefont
  {Volovik}},\ }\href {\doibase 10.1103/RevModPhys.59.533} {\bibfield
  {journal} {\bibinfo  {journal} {Rev. Mod. Phys.}\ }\textbf {\bibinfo {volume}
  {59}},\ \bibinfo {pages} {533} (\bibinfo {year} {1987})}\BibitemShut
  {NoStop}%
\bibitem [{\citenamefont {Nieh}\ \emph {et~al.}(1995)\citenamefont {Nieh},
  \citenamefont {Su},\ and\ \citenamefont {Zhao}}]{Nieh_95}%
  \BibitemOpen
  \bibfield  {author} {\bibinfo {author} {\bibfnamefont {H.~T.}\ \bibnamefont
  {Nieh}}, \bibinfo {author} {\bibfnamefont {G.}~\bibnamefont {Su}}, \ and\
  \bibinfo {author} {\bibfnamefont {B.-H.}\ \bibnamefont {Zhao}},\ }\href
  {\doibase 10.1103/PhysRevB.51.3760} {\bibfield  {journal} {\bibinfo
  {journal} {Phys. Rev. B}\ }\textbf {\bibinfo {volume} {51}},\ \bibinfo
  {pages} {3760} (\bibinfo {year} {1995})}\BibitemShut {NoStop}%
\bibitem [{\citenamefont {Yang}(1962)}]{Yang_1962}%
  \BibitemOpen
  \bibfield  {author} {\bibinfo {author} {\bibfnamefont {C.~N.}\ \bibnamefont
  {Yang}},\ }\href {\doibase 10.1103/RevModPhys.34.694} {\bibfield  {journal}
  {\bibinfo  {journal} {Rev. Mod. Phys.}\ }\textbf {\bibinfo {volume} {34}},\
  \bibinfo {pages} {694} (\bibinfo {year} {1962})}\BibitemShut {NoStop}%
\bibitem [{\citenamefont {Rampp}\ and\ \citenamefont
  {Schmalian}(2022)}]{Rampp_2022}%
  \BibitemOpen
  \bibfield  {author} {\bibinfo {author} {\bibfnamefont {M.~A.}\ \bibnamefont
  {Rampp}}\ and\ \bibinfo {author} {\bibfnamefont {J.}~\bibnamefont
  {Schmalian}},\ }\href {\doibase 10.1088/2399-6528/ac7033} {\bibfield
  {journal} {\bibinfo  {journal} {Journal of Physics Communications}\ }\textbf
  {\bibinfo {volume} {6}},\ \bibinfo {pages} {055013} (\bibinfo {year}
  {2022})}\BibitemShut {NoStop}%
\bibitem [{\citenamefont {Babaev}(2002)}]{Babaev_2002}%
  \BibitemOpen
  \bibfield  {author} {\bibinfo {author} {\bibfnamefont {E.}~\bibnamefont
  {Babaev}},\ }\href {https://doi.org/10.1103%2Fphysrevlett.89.067001}
  {\bibfield  {journal} {\bibinfo  {journal} {Physical Review Letters}\
  }\textbf {\bibinfo {volume} {89}} (\bibinfo {year} {2002})}\BibitemShut
  {NoStop}%
\bibitem [{\citenamefont {S\'a~de Melo}(1996)}]{Melo_96}%
  \BibitemOpen
  \bibfield  {author} {\bibinfo {author} {\bibfnamefont {C.~A.~R.}\
  \bibnamefont {S\'a~de Melo}},\ }\href {\doibase 10.1103/PhysRevB.53.R6010}
  {\bibfield  {journal} {\bibinfo  {journal} {Phys. Rev. B}\ }\textbf {\bibinfo
  {volume} {53}},\ \bibinfo {pages} {R6010} (\bibinfo {year}
  {1996})}\BibitemShut {NoStop}%
\bibitem [{\citenamefont {Schwartz}\ and\ \citenamefont
  {Cooper}(1965)}]{Scwartz_65}%
  \BibitemOpen
  \bibfield  {author} {\bibinfo {author} {\bibfnamefont {B.~B.}\ \bibnamefont
  {Schwartz}}\ and\ \bibinfo {author} {\bibfnamefont {L.~N.}\ \bibnamefont
  {Cooper}},\ }\href {\doibase 10.1103/PhysRev.137.A829} {\bibfield  {journal}
  {\bibinfo  {journal} {Phys. Rev.}\ }\textbf {\bibinfo {volume} {137}},\
  \bibinfo {pages} {A829} (\bibinfo {year} {1965})}\BibitemShut {NoStop}%
\bibitem [{\citenamefont {Schwartz}\ and\ \citenamefont
  {Cooper}(1964)}]{Schwartz_64}%
  \BibitemOpen
  \bibfield  {author} {\bibinfo {author} {\bibfnamefont {B.~B.}\ \bibnamefont
  {Schwartz}}\ and\ \bibinfo {author} {\bibfnamefont {L.~N.}\ \bibnamefont
  {Cooper}},\ }\href {\doibase 10.1103/RevModPhys.36.280} {\bibfield  {journal}
  {\bibinfo  {journal} {Rev. Mod. Phys.}\ }\textbf {\bibinfo {volume} {36}},\
  \bibinfo {pages} {280} (\bibinfo {year} {1964})}\BibitemShut {NoStop}%
\bibitem [{\citenamefont {Erlandsson}\ \emph {et~al.}(2022)\citenamefont
  {Erlandsson}, \citenamefont {Sabonis}, \citenamefont {Kringhøj},
  \citenamefont {Larsen}, \citenamefont {Krogstrup}, \citenamefont
  {Petersson},\ and\ \citenamefont {Marcus}}]{FS_Marcus22}%
  \BibitemOpen
  \bibfield  {author} {\bibinfo {author} {\bibfnamefont {O.}~\bibnamefont
  {Erlandsson}}, \bibinfo {author} {\bibfnamefont {D.}~\bibnamefont {Sabonis}},
  \bibinfo {author} {\bibfnamefont {A.}~\bibnamefont {Kringhøj}}, \bibinfo
  {author} {\bibfnamefont {T.~W.}\ \bibnamefont {Larsen}}, \bibinfo {author}
  {\bibfnamefont {P.}~\bibnamefont {Krogstrup}}, \bibinfo {author}
  {\bibfnamefont {K.~D.}\ \bibnamefont {Petersson}}, \ and\ \bibinfo {author}
  {\bibfnamefont {C.~M.}\ \bibnamefont {Marcus}},\ }\href {\doibase
  10.48550/ARXIV.2202.05974} {\  (\bibinfo {year} {2022}),\
  10.48550/ARXIV.2202.05974}\BibitemShut {NoStop}%
\bibitem [{\citenamefont {Sabonis}\ \emph {et~al.}(2020)\citenamefont
  {Sabonis}, \citenamefont {Erlandsson}, \citenamefont {Kringh\o{}j},
  \citenamefont {van Heck}, \citenamefont {Larsen}, \citenamefont {Petkovic},
  \citenamefont {Krogstrup}, \citenamefont {Petersson},\ and\ \citenamefont
  {Marcus}}]{FS_Marcus20}%
  \BibitemOpen
  \bibfield  {author} {\bibinfo {author} {\bibfnamefont {D.}~\bibnamefont
  {Sabonis}}, \bibinfo {author} {\bibfnamefont {O.}~\bibnamefont {Erlandsson}},
  \bibinfo {author} {\bibfnamefont {A.}~\bibnamefont {Kringh\o{}j}}, \bibinfo
  {author} {\bibfnamefont {B.}~\bibnamefont {van Heck}}, \bibinfo {author}
  {\bibfnamefont {T.~W.}\ \bibnamefont {Larsen}}, \bibinfo {author}
  {\bibfnamefont {I.}~\bibnamefont {Petkovic}}, \bibinfo {author}
  {\bibfnamefont {P.}~\bibnamefont {Krogstrup}}, \bibinfo {author}
  {\bibfnamefont {K.~D.}\ \bibnamefont {Petersson}}, \ and\ \bibinfo {author}
  {\bibfnamefont {C.~M.}\ \bibnamefont {Marcus}},\ }\href {\doibase
  10.1103/PhysRevLett.125.156804} {\bibfield  {journal} {\bibinfo  {journal}
  {Phys. Rev. Lett.}\ }\textbf {\bibinfo {volume} {125}},\ \bibinfo {pages}
  {156804} (\bibinfo {year} {2020})}\BibitemShut {NoStop}%
\bibitem [{\citenamefont {Zhu}(2016)}]{Jianxin}%
  \BibitemOpen
  \bibfield  {author} {\bibinfo {author} {\bibfnamefont {J.-X.}\ \bibnamefont
  {Zhu}},\ }\href@noop {} {\emph {\bibinfo {title} {Bogoliubov de Gennes
  Methods and its Applications}}}\ (\bibinfo  {publisher} {Springer},\ \bibinfo
  {year} {2016})\BibitemShut {NoStop}%
\bibitem [{\citenamefont {Bogoliubov}(1958)}]{Bogoliubov:1958km}%
  \BibitemOpen
  \bibfield  {author} {\bibinfo {author} {\bibfnamefont {N.~N.}\ \bibnamefont
  {Bogoliubov}},\ }\href {\doibase 10.1007/BF02745585} {\bibfield  {journal}
  {\bibinfo  {journal} {Nuovo Cim.}\ }\textbf {\bibinfo {volume} {7}},\
  \bibinfo {pages} {794–805} (\bibinfo {year} {1958})}\BibitemShut {NoStop}%
\bibitem [{\citenamefont {San-Jose}\ \emph {et~al.}(2022)\citenamefont
  {San-Jose}, \citenamefont {Payá}, \citenamefont {Marcus}, \citenamefont
  {Vaitiekėnas},\ and\ \citenamefont {Prada}}]{SanJose_2022}%
  \BibitemOpen
  \bibfield  {author} {\bibinfo {author} {\bibfnamefont {P.}~\bibnamefont
  {San-Jose}}, \bibinfo {author} {\bibfnamefont {C.}~\bibnamefont {Payá}},
  \bibinfo {author} {\bibfnamefont {C.~M.}\ \bibnamefont {Marcus}}, \bibinfo
  {author} {\bibfnamefont {S.}~\bibnamefont {Vaitiekėnas}}, \ and\ \bibinfo
  {author} {\bibfnamefont {E.}~\bibnamefont {Prada}},\ }\href {\doibase
  10.48550/ARXIV.2207.07606} {\  (\bibinfo {year} {2022}),\
  10.48550/ARXIV.2207.07606}\BibitemShut {NoStop}%
\bibitem [{\citenamefont {Ginzburg}\ and\ \citenamefont {Landau}(1950)}]{GL}%
  \BibitemOpen
  \bibfield  {author} {\bibinfo {author} {\bibfnamefont {V.~L.}\ \bibnamefont
  {Ginzburg}}\ and\ \bibinfo {author} {\bibfnamefont {L.~D.}\ \bibnamefont
  {Landau}},\ }\href@noop {} {\bibfield  {journal} {\bibinfo  {journal} {Zh.
  Eksp. Teor. Fiz.}\ }\textbf {\bibinfo {volume} {20}},\ \bibinfo {pages}
  {1064–1082} (\bibinfo {year} {1950})}\BibitemShut {NoStop}%
\bibitem [{\citenamefont {Fetter}\ and\ \citenamefont
  {Walecka}(2003)}]{FetterW}%
  \BibitemOpen
  \bibfield  {author} {\bibinfo {author} {\bibfnamefont {A.~L.}\ \bibnamefont
  {Fetter}}\ and\ \bibinfo {author} {\bibfnamefont {J.~D.}\ \bibnamefont
  {Walecka}},\ }\href@noop {} {\emph {\bibinfo {title} {Quantum Theory of
  Many-Particle Systems}}}\ (\bibinfo  {publisher} {New York: Dover.},\
  \bibinfo {year} {2003})\BibitemShut {NoStop}%
\bibitem [{\citenamefont {Tinkham}(2003)}]{Tinkham}%
  \BibitemOpen
  \bibfield  {author} {\bibinfo {author} {\bibfnamefont {M.}~\bibnamefont
  {Tinkham}},\ }\href@noop {} {\emph {\bibinfo {title} {Introduction to
  Superconductivity}}}\ (\bibinfo  {publisher} {Dover Publications, Inc.
  Mineola, New York.},\ \bibinfo {year} {2003})\BibitemShut {NoStop}%
\bibitem [{\citenamefont {Golubovi\ifmmode~\acute{c}\else \'{c}\fi{}}\ \emph
  {et~al.}(2003)\citenamefont {Golubovi\ifmmode~\acute{c}\else \'{c}\fi{}},
  \citenamefont {Pogosov}, \citenamefont {Morelle},\ and\ \citenamefont
  {Moshchalkov}}]{Golubov03}%
  \BibitemOpen
  \bibfield  {author} {\bibinfo {author} {\bibfnamefont {D.~S.}\ \bibnamefont
  {Golubovi\ifmmode~\acute{c}\else \'{c}\fi{}}}, \bibinfo {author}
  {\bibfnamefont {W.~V.}\ \bibnamefont {Pogosov}}, \bibinfo {author}
  {\bibfnamefont {M.}~\bibnamefont {Morelle}}, \ and\ \bibinfo {author}
  {\bibfnamefont {V.~V.}\ \bibnamefont {Moshchalkov}},\ }\href {\doibase
  10.1103/PhysRevB.68.172503} {\bibfield  {journal} {\bibinfo  {journal} {Phys.
  Rev. B}\ }\textbf {\bibinfo {volume} {68}},\ \bibinfo {pages} {172503}
  (\bibinfo {year} {2003})}\BibitemShut {NoStop}%
\bibitem [{\citenamefont {Eremko}\ \emph {et~al.}(2018)\citenamefont {Eremko},
  \citenamefont {Brizhik},\ and\ \citenamefont {Loktev}}]{SOIcorr}%
  \BibitemOpen
  \bibfield  {author} {\bibinfo {author} {\bibfnamefont {A.~A.}\ \bibnamefont
  {Eremko}}, \bibinfo {author} {\bibfnamefont {L.~S.}\ \bibnamefont {Brizhik}},
  \ and\ \bibinfo {author} {\bibfnamefont {V.~M.}\ \bibnamefont {Loktev}},\
  }\href {\doibase 10.1063/1.5037561} {\bibfield  {journal} {\bibinfo
  {journal} {Low Temperature Physics}\ }\textbf {\bibinfo {volume} {44}},\
  \bibinfo {pages} {573–583} (\bibinfo {year} {2018})}\BibitemShut {NoStop}%
\bibitem [{\citenamefont {Berche}\ \emph {et~al.}(2012)\citenamefont {Berche},
  \citenamefont {Medina},\ and\ \citenamefont {L{\'{o}}pez}}]{Berche_2012}%
  \BibitemOpen
  \bibfield  {author} {\bibinfo {author} {\bibfnamefont {B.}~\bibnamefont
  {Berche}}, \bibinfo {author} {\bibfnamefont {E.}~\bibnamefont {Medina}}, \
  and\ \bibinfo {author} {\bibfnamefont {A.}~\bibnamefont {L{\'{o}}pez}},\
  }\href {\doibase 10.1209/0295-5075/97/67007} {\bibfield  {journal} {\bibinfo
  {journal} {{EPL} (Europhysics Letters)}\ }\textbf {\bibinfo {volume} {97}},\
  \bibinfo {pages} {67007} (\bibinfo {year} {2012})}\BibitemShut {NoStop}%
\bibitem [{\citenamefont {Gurtler}\ and\ \citenamefont
  {Hestenes}(1975)}]{Hestenes_consistency}%
  \BibitemOpen
  \bibfield  {author} {\bibinfo {author} {\bibfnamefont {R.}~\bibnamefont
  {Gurtler}}\ and\ \bibinfo {author} {\bibfnamefont {D.}~\bibnamefont
  {Hestenes}},\ }\href {\doibase 10.1063/1.522555} {\bibfield  {journal}
  {\bibinfo  {journal} {Journal of Mathematical Physics}\ }\textbf {\bibinfo
  {volume} {16}},\ \bibinfo {pages} {573} (\bibinfo {year} {1975})}\BibitemShut
  {NoStop}%
\bibitem [{\citenamefont {Odom}\ \emph {et~al.}(2006)\citenamefont {Odom},
  \citenamefont {Hanneke}, \citenamefont {D'Urso},\ and\ \citenamefont
  {Gabrielse}}]{Odom_06}%
  \BibitemOpen
  \bibfield  {author} {\bibinfo {author} {\bibfnamefont {B.}~\bibnamefont
  {Odom}}, \bibinfo {author} {\bibfnamefont {D.}~\bibnamefont {Hanneke}},
  \bibinfo {author} {\bibfnamefont {B.}~\bibnamefont {D'Urso}}, \ and\ \bibinfo
  {author} {\bibfnamefont {G.}~\bibnamefont {Gabrielse}},\ }\href {\doibase
  10.1103/PhysRevLett.97.030801} {\bibfield  {journal} {\bibinfo  {journal}
  {Phys. Rev. Lett.}\ }\textbf {\bibinfo {volume} {97}},\ \bibinfo {pages}
  {030801} (\bibinfo {year} {2006})}\BibitemShut {NoStop}%
\bibitem [{\citenamefont {Nilsson}\ \emph {et~al.}(2009)\citenamefont
  {Nilsson}, \citenamefont {Caroff}, \citenamefont {Thelander}, \citenamefont
  {Larsson}, \citenamefont {Wagner}, \citenamefont {Wernersson}, \citenamefont
  {Samuelson},\ and\ \citenamefont {Xu}}]{Nilsson2009}%
  \BibitemOpen
  \bibfield  {author} {\bibinfo {author} {\bibfnamefont {H.~A.}\ \bibnamefont
  {Nilsson}}, \bibinfo {author} {\bibfnamefont {P.}~\bibnamefont {Caroff}},
  \bibinfo {author} {\bibfnamefont {C.}~\bibnamefont {Thelander}}, \bibinfo
  {author} {\bibfnamefont {M.}~\bibnamefont {Larsson}}, \bibinfo {author}
  {\bibfnamefont {J.~B.}\ \bibnamefont {Wagner}}, \bibinfo {author}
  {\bibfnamefont {L.-E.}\ \bibnamefont {Wernersson}}, \bibinfo {author}
  {\bibfnamefont {L.}~\bibnamefont {Samuelson}}, \ and\ \bibinfo {author}
  {\bibfnamefont {H.~Q.}\ \bibnamefont {Xu}},\ }\href {\doibase
  10.1021/nl901333a} {\bibfield  {journal} {\bibinfo  {journal} {Nano Letters}\
  }\textbf {\bibinfo {volume} {9}},\ \bibinfo {pages} {3151} (\bibinfo {year}
  {2009})}\BibitemShut {NoStop}%
\bibitem [{\citenamefont {{Bychkov}}\ and\ \citenamefont
  {{Rashba}}(1984)}]{Rashba84}%
  \BibitemOpen
  \bibfield  {author} {\bibinfo {author} {\bibfnamefont {Y.~A.}\ \bibnamefont
  {{Bychkov}}}\ and\ \bibinfo {author} {\bibfnamefont {{\'E}.~I.}\ \bibnamefont
  {{Rashba}}},\ }\href@noop {} {\bibfield  {journal} {\bibinfo  {journal}
  {Soviet Journal of Experimental and Theoretical Physics Letters}\ }\textbf
  {\bibinfo {volume} {39}},\ \bibinfo {pages} {78} (\bibinfo {year}
  {1984})}\BibitemShut {NoStop}%
\bibitem [{\citenamefont {Dresselhaus}(1955)}]{Dresselhaus55}%
  \BibitemOpen
  \bibfield  {author} {\bibinfo {author} {\bibfnamefont {G.}~\bibnamefont
  {Dresselhaus}},\ }\href {\doibase 10.1103/PhysRev.100.580} {\bibfield
  {journal} {\bibinfo  {journal} {Phys. Rev.}\ }\textbf {\bibinfo {volume}
  {100}},\ \bibinfo {pages} {580} (\bibinfo {year} {1955})}\BibitemShut
  {NoStop}%
\bibitem [{\citenamefont {Bringer}\ and\ \citenamefont
  {Schäpers}(2011)}]{Bringer2011}%
  \BibitemOpen
  \bibfield  {author} {\bibinfo {author} {\bibfnamefont {A.}~\bibnamefont
  {Bringer}}\ and\ \bibinfo {author} {\bibfnamefont {T.}~\bibnamefont
  {Schäpers}},\ }\href {\doibase 10.1103/PhysRevB.83.115305} {\bibfield
  {journal} {\bibinfo  {journal} {Phys. Rev. B}\ }\textbf {\bibinfo {volume}
  {83}},\ \bibinfo {pages} {115305} (\bibinfo {year} {2011})}\BibitemShut
  {NoStop}%
\bibitem [{\citenamefont {Nambu}(1960)}]{Nambu_1960}%
  \BibitemOpen
  \bibfield  {author} {\bibinfo {author} {\bibfnamefont {Y.}~\bibnamefont
  {Nambu}},\ }\href {\doibase 10.1103/PhysRev.117.648} {\bibfield  {journal}
  {\bibinfo  {journal} {Phys. Rev.}\ }\textbf {\bibinfo {volume} {117}},\
  \bibinfo {pages} {648} (\bibinfo {year} {1960})}\BibitemShut {NoStop}%
\bibitem [{\citenamefont {Stanescu}\ and\ \citenamefont {{Das
  Sarma}}(2017)}]{TudorSarma}%
  \BibitemOpen
  \bibfield  {author} {\bibinfo {author} {\bibfnamefont {T.~D.}\ \bibnamefont
  {Stanescu}}\ and\ \bibinfo {author} {\bibfnamefont {S.}~\bibnamefont {{Das
  Sarma}}},\ }\href {\doibase 10.1103/PhysRevB.96.014510} {\bibfield  {journal}
  {\bibinfo  {journal} {Phys. Rev. B}\ }\textbf {\bibinfo {volume} {96}},\
  \bibinfo {pages} {014510} (\bibinfo {year} {2017})}\BibitemShut {NoStop}%
\bibitem [{\citenamefont {Gül}\ \emph {et~al.}(2014)\citenamefont {Gül},
  \citenamefont {Gunel}, \citenamefont {Lüth}, \citenamefont {Rieger},
  \citenamefont {Wenz}, \citenamefont {Haas}, \citenamefont {Lepsa},
  \citenamefont {Panaitov}, \citenamefont {Grützmacher},\ and\ \citenamefont
  {Schaepers}}]{GulShapers_2014}%
  \BibitemOpen
  \bibfield  {author} {\bibinfo {author} {\bibfnamefont {O.}~\bibnamefont
  {Gül}}, \bibinfo {author} {\bibfnamefont {Y.}~\bibnamefont {Gunel}},
  \bibinfo {author} {\bibfnamefont {H.}~\bibnamefont {Lüth}}, \bibinfo
  {author} {\bibfnamefont {T.}~\bibnamefont {Rieger}}, \bibinfo {author}
  {\bibfnamefont {T.}~\bibnamefont {Wenz}}, \bibinfo {author} {\bibfnamefont
  {F.}~\bibnamefont {Haas}}, \bibinfo {author} {\bibfnamefont {M.}~\bibnamefont
  {Lepsa}}, \bibinfo {author} {\bibfnamefont {G.}~\bibnamefont {Panaitov}},
  \bibinfo {author} {\bibfnamefont {D.}~\bibnamefont {Grützmacher}}, \ and\
  \bibinfo {author} {\bibfnamefont {T.}~\bibnamefont {Schaepers}},\ }\href
  {\doibase 10.1021/nl502598s} {\bibfield  {journal} {\bibinfo  {journal} {Nano
  letters}\ }\textbf {\bibinfo {volume} {14}} (\bibinfo {year} {2014}),\
  10.1021/nl502598s}\BibitemShut {NoStop}%
\bibitem [{\citenamefont {Önder Gül}\ \emph {et~al.}(2017)\citenamefont
  {Önder Gül}, \citenamefont {Zhang}, \citenamefont {de~Vries}, \citenamefont
  {van Veen}, \citenamefont {Zuo}, \citenamefont {Mourik}, \citenamefont
  {Conesa-Boj}, \citenamefont {Nowak}, \citenamefont {van Woerkom},
  \citenamefont {Quintero-P{\'{e} }rez}, \citenamefont {Cassidy}, \citenamefont
  {Geresdi}, \citenamefont {Koelling}, \citenamefont {Car}, \citenamefont
  {Plissard}, \citenamefont {Bakkers},\ and\ \citenamefont
  {Kouwenhoven}}]{Gul_2017}%
  \BibitemOpen
  \bibfield  {author} {\bibinfo {author} {\bibnamefont {Önder Gül}}, \bibinfo
  {author} {\bibfnamefont {H.}~\bibnamefont {Zhang}}, \bibinfo {author}
  {\bibfnamefont {F.~K.}\ \bibnamefont {de~Vries}}, \bibinfo {author}
  {\bibfnamefont {J.}~\bibnamefont {van Veen}}, \bibinfo {author}
  {\bibfnamefont {K.}~\bibnamefont {Zuo}}, \bibinfo {author} {\bibfnamefont
  {V.}~\bibnamefont {Mourik}}, \bibinfo {author} {\bibfnamefont
  {S.}~\bibnamefont {Conesa-Boj}}, \bibinfo {author} {\bibfnamefont {M.~P.}\
  \bibnamefont {Nowak}}, \bibinfo {author} {\bibfnamefont {D.~J.}\ \bibnamefont
  {van Woerkom}}, \bibinfo {author} {\bibfnamefont {M.}~\bibnamefont
  {Quintero-P{\'{e} }rez}}, \bibinfo {author} {\bibfnamefont {M.~C.}\
  \bibnamefont {Cassidy}}, \bibinfo {author} {\bibfnamefont {A.}~\bibnamefont
  {Geresdi}}, \bibinfo {author} {\bibfnamefont {S.}~\bibnamefont {Koelling}},
  \bibinfo {author} {\bibfnamefont {D.}~\bibnamefont {Car}}, \bibinfo {author}
  {\bibfnamefont {S.~R.}\ \bibnamefont {Plissard}}, \bibinfo {author}
  {\bibfnamefont {E.~P. A.~M.}\ \bibnamefont {Bakkers}}, \ and\ \bibinfo
  {author} {\bibfnamefont {L.~P.}\ \bibnamefont {Kouwenhoven}},\ }\href
  {\doibase 10.1021/acs.nanolett.7b00540} {\bibfield  {journal} {\bibinfo
  {journal} {Nano Letters}\ }\textbf {\bibinfo {volume} {17}},\ \bibinfo
  {pages} {2690} (\bibinfo {year} {2017})}\BibitemShut {NoStop}%
\bibitem [{\citenamefont {Wu}\ \emph {et~al.}(1992)\citenamefont {Wu},
  \citenamefont {Sprung},\ and\ \citenamefont {Martorell}}]{Wu_92}%
  \BibitemOpen
  \bibfield  {author} {\bibinfo {author} {\bibfnamefont {H.}~\bibnamefont
  {Wu}}, \bibinfo {author} {\bibfnamefont {D.~W.~L.}\ \bibnamefont {Sprung}}, \
  and\ \bibinfo {author} {\bibfnamefont {J.}~\bibnamefont {Martorell}},\ }\href
  {\doibase http://dx.doi.org/10.1063/1.352176} {\bibfield  {journal} {\bibinfo
   {journal} {Journal of Applied Physics}\ }\textbf {\bibinfo {volume} {72}},\
  \bibinfo {pages} {151} (\bibinfo {year} {1992})}\BibitemShut {NoStop}%
\bibitem [{\citenamefont {Ballester}\ \emph {et~al.}(2014)\citenamefont
  {Ballester}, \citenamefont {Segarra}, \citenamefont {Bertoni},\ and\
  \citenamefont {Planelles}}]{Ballester_2013}%
  \BibitemOpen
  \bibfield  {author} {\bibinfo {author} {\bibfnamefont {A.}~\bibnamefont
  {Ballester}}, \bibinfo {author} {\bibfnamefont {C.}~\bibnamefont {Segarra}},
  \bibinfo {author} {\bibfnamefont {A.}~\bibnamefont {Bertoni}}, \ and\
  \bibinfo {author} {\bibfnamefont {J.}~\bibnamefont {Planelles}},\ }\href
  {\doibase 10.1209/0295-5075/104/67004} {\bibfield  {journal} {\bibinfo
  {journal} {Europhysics Letters}\ }\textbf {\bibinfo {volume} {104}},\
  \bibinfo {pages} {67004} (\bibinfo {year} {2014})}\BibitemShut {NoStop}%
\bibitem [{\citenamefont {Urbaneja~Torres}\ \emph {et~al.}(2018)\citenamefont
  {Urbaneja~Torres}, \citenamefont {Sitek}, \citenamefont {Erlingsson},
  \citenamefont {Thorgilsson}, \citenamefont {Gudmundsson},\ and\ \citenamefont
  {Manolescu}}]{Torres_2018}%
  \BibitemOpen
  \bibfield  {author} {\bibinfo {author} {\bibfnamefont {M.}~\bibnamefont
  {Urbaneja~Torres}}, \bibinfo {author} {\bibfnamefont {A.}~\bibnamefont
  {Sitek}}, \bibinfo {author} {\bibfnamefont {S.~I.}\ \bibnamefont
  {Erlingsson}}, \bibinfo {author} {\bibfnamefont {G.}~\bibnamefont
  {Thorgilsson}}, \bibinfo {author} {\bibfnamefont {V.}~\bibnamefont
  {Gudmundsson}}, \ and\ \bibinfo {author} {\bibfnamefont {A.}~\bibnamefont
  {Manolescu}},\ }\href {\doibase 10.1103/PhysRevB.98.085419} {\bibfield
  {journal} {\bibinfo  {journal} {Phys. Rev. B}\ }\textbf {\bibinfo {volume}
  {98}},\ \bibinfo {pages} {085419} (\bibinfo {year} {2018})}\BibitemShut
  {NoStop}%
\bibitem [{\citenamefont {Niemelä}\ \emph {et~al.}(1996)\citenamefont
  {Niemelä}, \citenamefont {Pietiläinen}, \citenamefont {Hyvönen},\ and\
  \citenamefont {Chakraborty}}]{Niemela_1996}%
  \BibitemOpen
  \bibfield  {author} {\bibinfo {author} {\bibfnamefont {K.}~\bibnamefont
  {Niemelä}}, \bibinfo {author} {\bibfnamefont {P.}~\bibnamefont
  {Pietiläinen}}, \bibinfo {author} {\bibfnamefont {P.}~\bibnamefont
  {Hyvönen}}, \ and\ \bibinfo {author} {\bibfnamefont {T.}~\bibnamefont
  {Chakraborty}},\ }\href {\doibase 10.1209/epl/i1996-00265-7} {\bibfield
  {journal} {\bibinfo  {journal} {Europhysics Letters}\ }\textbf {\bibinfo
  {volume} {36}},\ \bibinfo {pages} {533} (\bibinfo {year} {1996})}\BibitemShut
  {NoStop}%
\end{thebibliography}%
\end{document}